\documentclass[%
 reprint,prl,
superscriptaddress,
showpacs,
 amsmath,amssymb,
 aps,
]{revtex4-1}

\usepackage[export]{adjustbox}

\pdfoutput=1
\usepackage{graphicx}
\usepackage{dcolumn}
\usepackage{bm}
\usepackage{hyperref}
\usepackage{verbatim} 
\usepackage{color}
\usepackage{pifont}
\usepackage{bbold}
\usepackage[normalem]{ulem}

\usepackage{xr}

\begin{document}

\title{Generic transport mechanisms for molecular traffic in cellular protrusions}

\author{Isabella R. Graf}%
\affiliation{%
Arnold-Sommerfeld-Center for Theoretical Physics and Center for NanoScience, \\
Department of Physics, Ludwig-Maximilians-Universit\"at M\"unchen, D--80333 Munich, Germany
}
\author{Erwin Frey}%
\altaffiliation{Correspondence please to frey@lmu.de.}
\affiliation{%
Arnold-Sommerfeld-Center for Theoretical Physics and Center for NanoScience, \\
Department of Physics, Ludwig-Maximilians-Universit\"at M\"unchen, D--80333 Munich, Germany
}
             
\begin{abstract}
Transport of molecular motors along protein filaments in a half-closed geometry is a common feature of biologically relevant processes in cellular protrusions. Using a lattice gas model we study how the interplay between active and diffusive transport and mass conservation leads to localised domain walls and tip localisation of the motors. 
We identify a mechanism for task sharing between the active motors (maintaining a gradient) and the diffusive motion (transport to the tip), which ensures that energy consumption is low and motor exchange mostly happens at the tip.
These features are attributed to strong nearest-neighbour correlations that lead to a strong reduction of active currents, which we calculate analytically using an exact moment-identity, and might prove useful for the understanding of correlations and active transport also in more elaborate systems.
\end{abstract}



\maketitle

Linear protrusions of cells, as for instance filopodia or stereocilia, perform multiple tasks in living organisms, ranging from cell migration and signal transduction to wound healing. They contain actin filaments cross-linked into bundles by actin-binding proteins~\cite{Revenu:2004aa,Mattila:2008aa,LesErickson01102003,Nambiar2010}, and molecular motors of the myosin family which interact with actin filaments and walk on them in a persistent, uni-directional fashion towards the tip of the protrusion~\cite{LesErickson01102003, Nambiar2010, Salles:2009aa, Hartman1627, Kerber3733, Kambara08122006, Bird26082014}. These motors play an important role in the biological function of protrusions~\cite{Revenu:2004aa,Nambiar2010,Salles:2009aa}. In particular, they are known to localise to the tips of filopodia and stereocilia, and are (jointly) responsible for length control~\cite{Mattila:2008aa, LesErickson01102003, Nambiar2010,Salles:2009aa,Hartman1627, Kerber3733, Kambara08122006, Bird26082014, Rzadzinska15032004, Manor2011167, Belyantseva25112003}.

Motivated by these observations, various models have been investigated. 
Some are detailed mathematical models addressing specific biological issues. These include the role of motor transport in shaping the concentration profile of G-actin at the base of protrusions~\cite{zhuravlev2}, the localisation of different proteins along stereocilia~\cite{NaozGov}, and the effect of myosin X on filopodial growth~\cite{1478-3975-11-1-016005}. Complementary, simplified conceptual models have been studied asking how the interplay between active and diffusive transport in open systems may lead to non-equilibrium phase transitions and ensuing steady states with interesting correlations and nontrivial density profiles~\cite{PhysRevLett.67.1882, lipowsky, PhysRevLett.90.086601, PhysRevE.70.046101, evans}. The latter are based on the totally asymmetric simple exclusion process (TASEP) \cite{BIP:BIP360060102, Spitzer1970}, a lattice-gas  model that, despite its simple structure, has become a paradigm for non-equilibrium dynamics \cite{blythe-evans:2007, Chou0034-4885-74-11-116601}.

Here we present and analyse a conceptual model capturing two basic properties of the motion of persistent, plus-end directed motors inside narrow, elongated cellular protrusions. First, there is an interplay between two genuinely different types of dynamics: \emph{directed (active) transport} with steric hindrance along actin filaments, and \emph{diffusive motion} in the cytoplasm. These are coupled by particle exchange between the filament and the cytoplasm. Second, the \emph{half-closed geometry} of cellular protrusions is special: At one end, the protrusions are connected to the cell body and thus to a reservoir, whereas everywhere else protein diffusion is confined by the cell membrane, so that mass conservation and resource limitation play an important role there. The combination  of mass-conservation (closure) on the one hand and the interplay of equilibrium (diffusion) and non-equilibrium (active transport) processes on the other hand is intrinsically interesting to study as closure in a system entails no-flux boundary conditions that oppose currents from active transport. Here, we want to examine the interplay of these mechanisms with the help of an abstract model that is motivated biologically but has a level of description that makes it possible to understand all the processes accounted for. We identify generic mechanisms based on correlations and non-equilibrium physics that could be of importance for biological systems as cellular protrusions but, inevitably, predictions for biological systems are qualitative.

\begin{figure}[t]
    \centering
    \includegraphics[width=\linewidth]{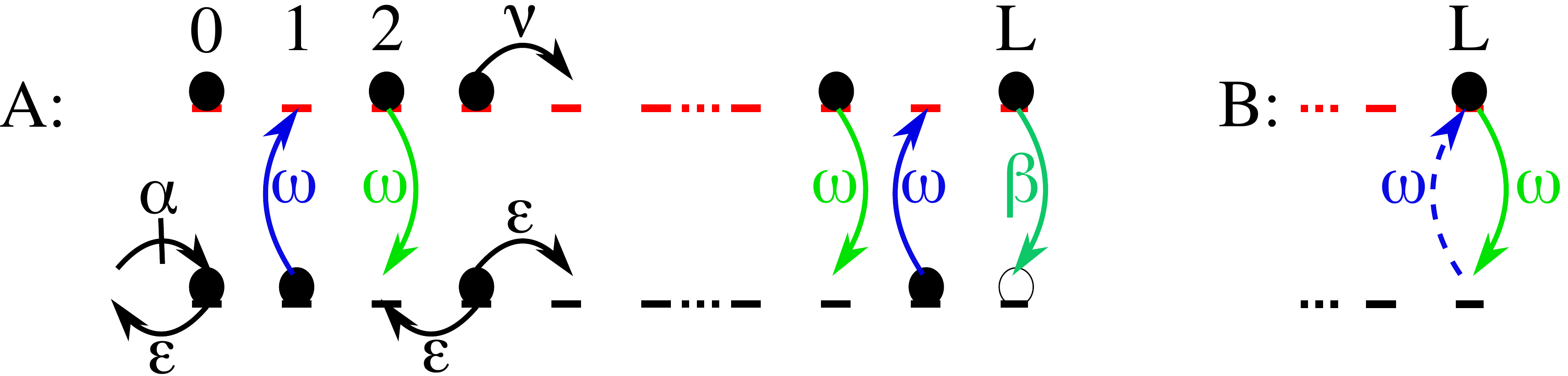}
    \caption{Illustration of the two-lane lattice-gas model comprised of a TASEP and a SSEP with hopping rates $\nu$ and $\epsilon$, respectively. The lanes are coupled via a symmetric attachment and detachment rate $\omega \,{\ll}\,  \nu$. In model A, detachment from the last site is given by $\beta \,{\gg}\, \omega$, while in model B exchange between the lanes is fully symmetric. At the tip of the systems mass conservation holds, and at  the base particles can enter the SSEP lane at rate $\alpha$ and exit at rate $\epsilon$.}
    \label{fig:model}
\end{figure}

Specifically, we consider a two-lane lattice-gas model in a half-closed geometry [Fig.\ \ref{fig:model}] in steady state, similar to Ref.\ \cite{mueller}.  One lane represents the actin filament and the second lane the cytoplasm. While there is a rich literature on the non-equilibrium dynamics of two-lane systems \cite{klumpp, Popkov2003, 0305-4470-37-42-005, 0295-5075-70-3-299, mueller, Pronina200612, 1367-2630-9-6-159, 1751-8121-40-10-004, Jiang2007247, Reichenbach2008, Wang2008457, Evans2009, 1742-5468-2010-06-P06002, evans, PhysRevE.83.031923, saha,  PhysRevE.89.022131, PhysRevE.89.042713, pinkoviezky, 1751-8121-49-9-095601}, very few of these studies address how the physics of non-equilibrium steady-states is affected by a half-closed geometry \cite{mueller, pinkoviezky}. We are interested in the limit where the actin filament and the cytoplasm are coupled weakly by attachment and detachment processes, while Ref.\ \cite{mueller} focuses on the strong coupling limit. Our analyses show that, due to the closure of the system at the tip, there is only one type of density profile, namely, domain walls (DW) separating a high- and a low-density region. The limit where the width of the high-density region becomes microscopically small (of the order of a few lattice sites) corresponds to tip localisation. Furthermore, correlations in such systems have not received much attention, and we want to investigate nearest-neighbour (NN) correlations on the filament. In a biological context, this is related to efficient transport on actin filaments  and to the significance of steric hindrance of motors there. We find that those correlations reach high values close to the DW, and the transport current along the filament is strongly reduced compared to its mean-field (MF) prediction. This suggests an important role for the cytoplasm, namely to transport the proteins to the tip.  Conversely, active transport effectively sets up and maintains a gradient of motor proteins.

Our model consists of two coupled sublattices [Fig.\  \ref{fig:model}], namely a TASEP and a SSEP (symmetric simple exclusion process) lane of $L+1$ sites $\,{\in}\, \{ 0,1,...,L \}$. The dynamics on lane $1$ (TASEP) are governed by a rate $\nu$ at which particles jump forward one site towards the right (tip) provided that the site in front of them is empty (exclusion). This corresponds to the directional motion of the motors on the filament that is oriented towards the tip. By convention, we  measure time in units of $\nu$ (i.e.\ set $\nu \,{=}\, 1$) and length in units of the system size (i.e.\ the lattice spacing is $a\,{=}\,\frac{1}{L}$ and the total length $L a \,{=}\,1$). In lane $2$ (SSEP) particles jump non-directionally between neighbouring sites, at  rate $\epsilon$ again respecting exclusion. This represents the diffusive motion in the cytoplasm that is taken to be effectively one-dimensional in the thin cylinder-like protrusions (the steady-state behaviour of a system with several lanes for diffusion arranged on a cylinder around the TASEP can be reduced to the steady-state behaviour of this model~\cite{Supplement}).  Particles enter or exit the system only at site $0$ of lane $2$ (base) but not at site 0 of lane 1. At rate $\alpha$ a particle is injected provided the site is empty and at rate $\epsilon$ a particle leaves the system. This reflects the exchange of motors between the protrusion base and the cell body. In the bulk both lanes are coupled via a rate $\omega$ at which particles jump from site $i \,{\in}\, \{ 0,1,...,L-1 \}$ of lane $1$ to site $i$ of lane $2$ or vice versa (each respecting exclusion).  Since the biochemistry at the tip is only poorly understood, at site $L$ we consider two extreme cases: In model A, particles jump from site $L$ of lane 1 to site $L$ of lane $2$ at rate $\beta \,{\gg}\, \omega$ (respecting exclusion) but not in the opposite direction. This describes a scenario where motors at the tip detach mainly due to the lack of a filament subunit in front of them. In model B, the exchange rates between the lanes at sites $L$ are the same as in the bulk. The comparison of both models stresses that seemingly minor changes in a non-equilibrium system may have a strong influence on the dynamics~\cite{Reese20140031}, and highlights the relevance of the biochemistry at the filament tip for the motor density profile. In the following, when considering the continuum limit $a\rightarrow 0$, we focus on the mesoscopic limit for $\omega$~\cite{PhysRevLett.90.086601, PhysRevE.70.046101}, i.e.\ we keep $\Omega \,{=}\, \omega/a$ fixed, thus ensuring the number of jumps between lanes is of the same order as that of entry or exit events (persistent motors) and implementing weak coupling between the diffusive and the directed motion.  For simplicity, we take the attachment and detachment rates to be equal. However, the qualitative results do not change for different attachment and detachment rates $\omega_A \neq \omega_D$ as long as both are still taken in the mesoscopic limit (not shown here).

\begin{figure}[t]
    \centering
    \includegraphics[trim={0cm 0cm 0cm 0cm},clip,width=\linewidth]{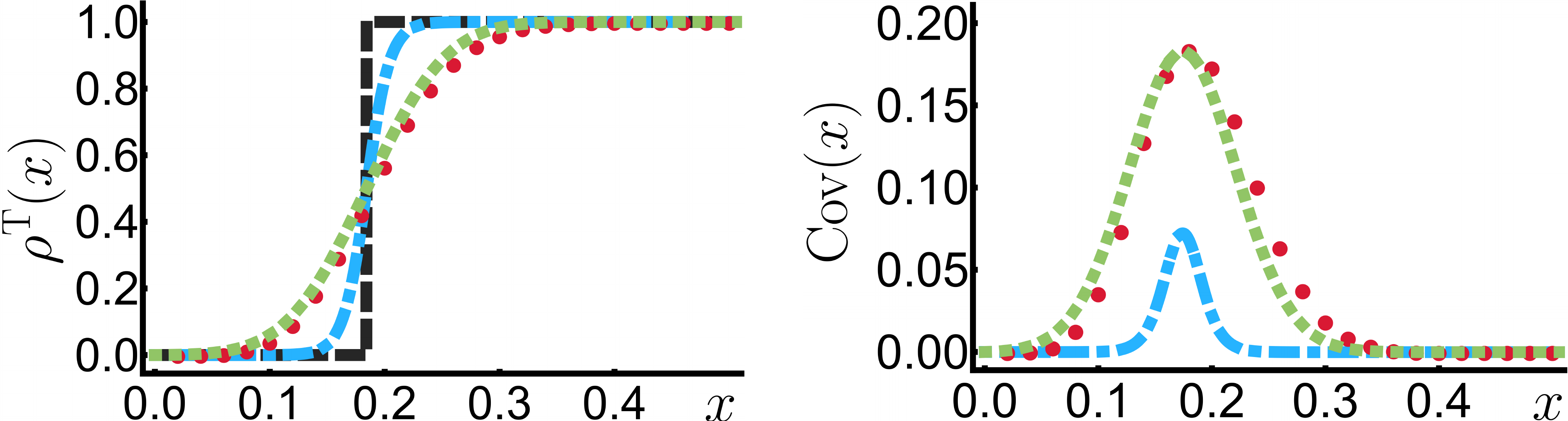}
    \caption{Representative steady-state DW profile $\rho^{\mathrm{T}} (x)$ (left) with covariances $\mathrm{Cov} (x)$ (right) on the TASEP lane are shown exemplarily for model A with $L \,{=}\, 50$, $\beta \,{=}\, 0.2$, $\Omega \,{=}\, 0.001$, $\alpha \,{=}\, 0.1$ and $\epsilon \,{=}\, 0.025$.
     The covariances are non-zero only in the vicinity of the DW. Using the fluctuation-corrected profile from DW theory, our predictions (dotted green curves) fit the simulation result (filled red circles) very well. If we use the refined MF profile (dot-dashed blue) instead of a step function (black dashed curve), the width of the DW and the strength of the covariances are strongly underestimated~\cite{Supplement}.}
    \label{fig:DWcor}
\end{figure}

A single TASEP exhibits three phases, namely a maximal current (MC), a low- (LD) and a high-density (HD) phase \cite{PhysRevLett.67.1882}. Moreover, on the phase boundary between the LD and HD phase the steady-state profile is given by a DW that performs a random walk. Due to the closure at the tip, we do not find a MC phase in our system~\cite{Supplement}, but instead observe \textit{localised} DWs~\cite{lipowsky, PhysRevLett.90.086601}. That is, the generic steady-state TASEP profile $\rho^{\mathrm{T}} (x)$ for both models is given by a localised DW separating low density at the base from high density at the tip [Fig.\ \ref{fig:DWcor}]. For generic parameters, the filament current is thus comparatively small, and restricted to a small part of the system. This might be beneficial from a biological point of view, since every motor step on the filament consumes ATP. The position of the DW depends on the model parameters, and is shifted towards the tip (base) for smaller (higher) values of $\sigma \,{:=}\, \frac{\alpha}{\alpha+\epsilon}$~\cite{Supplement}. This ratio can be thought of as the motor density in the cell body, and thus the value for the cytoplasmic density at the protrusion base. The width of the DW decreases with increasing $L$ and with increasing distance from the tip~\cite{Supplement}. Generically, when describing the DW as a random walker (see below), we observe that its motion is mainly confined to a small part of the system. 
Hence we assume for the moment that the DW is fully localised and adopt a step-function ansatz for $\rho^{\mathrm{T}} (x)$ that holds to lowest order in $a$: $\rho^{\mathrm{T}} (x) \,{=}\, 0$ for $x \,{\in}\, {[0,z}[ $, and $\rho^{\mathrm{T}} (x) \,{=}\, 1$ for $x \,{\in}\, ]z,1]$. First, we determine the position $z$ of the DW to find out whether tip localisation (i.e.\  $z \,{\approx}\, 1$) occurs for certain parameter ranges or not. For this purpose we derive a mass-balance equation relating the total average occupancies of the TASEP ($\mathrm{T}$) and the SSEP ($\mathrm{S}$). We denote by $n^{\mu}_i$ the occupation number on site $i \,{\in}\, \{0, \ldots, L\}$ of lane $\mu \,{\in}\, \{\mathrm{T},\mathrm{S}\}$, i.e.\ we write $n^{\mu}_i \,{=}\, 0$ if site $i$ of lane $\mu$ is empty and $n^{\mu}_i \,{=}\, 1$ if it is occupied. Since $n^{\mu}_i \,{\in}\, \{0,1\}$, we have $\langle n^{\mu}_i \rangle \,{=}\, \mathrm{Prob} \{n^{\mu}_i \,{=}\, 1 \}$; ensemble averages are denoted by $\langle \  \rangle$.  The steady-state condition in the bulk corresponds to a flux balance~\cite{Supplement}:
$
(\rho^{\mathrm{T}}_{i-1} \,{-}\,  f_{i-1} ) \,{-}\,  ( \rho^{\mathrm{T}}_i\,{-}\, f_i ) \,{=}\,    \omega (\rho^{\mathrm{T}}_i \,{-}\, \rho^{\mathrm{S}}_i)
$
with the correlator $f_i \,{=}\, \langle n^{\mathrm{T}}_i n^{\mathrm{T}}_{i+1} \rangle$ and the average occupancy  $\rho^{\mu}_i \,{=}\, \langle n^{\mu}_i \rangle$. The difference of the TASEP currents, from site $i\,{-}\, 1$ to site $i$ and from site $i$ to site $i\,{+}\, 1$, (left-hand side) must equal the current between sites $i$ of the TASEP and SSEP, $ \omega (\langle n^{\mathrm{T}}_i (1-n^{\mathrm{S}}_i) \rangle \,{-}\, \langle n^{\mathrm{S}}_i (1-n^{\mathrm{T}}_i) \rangle) $, (right-hand side). It follows that
$
f_i \,{=}\, f_0 \,{+}\, \rho^{\mathrm{T}}_i \,{-}\,\rho^{\mathrm{T}}_0 \,{+}\, \omega \sum_{j=1}^i (\rho^{\mathrm{T}}_j \,{-}\, \rho^{\mathrm{S}}_j)
$.
At the base, there is no direct flux from the cell body into the filament; so the boundary condition is $f_0\,{=}\,\rho^{\mathrm{T}}_0 + \omega (\rho^{\mathrm{T}}_0 - \rho^{\mathrm{S}}_0)$ and we find the following exact \textit{moment-identity}
\begin{align}
f_i = \rho^{\mathrm{T}}_i +\omega \sum_{j=0}^i (\rho^{\mathrm{T}}_j - \rho^{\mathrm{S}}_j) \, . 
\label{fi}
\end{align}
The systems are both closed at the tip but, due to the different attachment and detachment behaviour, the boundary conditions read $f_{L-1} \, {=} \, \rho^{\mathrm{T}}_{L-1} \,{-}\, \beta  \langle n^{\mathrm{T}}_L (1{-}n^{\mathrm{S}}_L) \rangle$, and $f_{L-1} \,{=}\, \rho^{\mathrm{T}}_{L-1} \,{+}\, \omega (\rho^{\mathrm{S}}_{L} \,{-}\, \rho^{\mathrm{T}}_L)$ for models A and B, respectively. Combining these with Eq.\ \eqref{fi}, the following exact \textit{mass-balance equations} can be derived for model A
\begin{align}
	\omega \sum_{j=0}^{L-1} \left( \rho^{\mathrm{T}}_j - \rho^{\mathrm{S}}_j \right) 
	= - \beta \left( \rho^{\mathrm{T}}_L - \left\langle n^{\mathrm{T}}_L n^{\mathrm{S}}_L \right\rangle \right) \, ,
\label{model1conservation} 
\end{align}
and similarly for model B: $\omega \sum_{j=0}^L (\rho^{\mathrm{T}}_j {-} \rho^{\mathrm{S}}_j) \,{=}\, 0$. These equations relate the average occupancy on the two lanes in such a way that the total influx into the TASEP lane equals the total outflux from it~\cite{Klumpp2007}. Interestingly, these equations reveal that a global detailed balance holds for the total exchange between the two lanes, rather than local detailed balance for any pair of sites (cf. adsorption equilibrium in Ref. \cite{mueller}). The moment-identity and the mass-balance equations are useful in two ways: (i) By using the DW ansatz, we are able to find an analytic formula for the DW position. (ii) The moment-identity enables us to express covariances with respect to NNs on the TASEP lane in terms of densities. 

\begin{figure}[t]
    \centering
    \includegraphics[trim={0cm 0cm 0cm 0cm},clip,width=\linewidth]{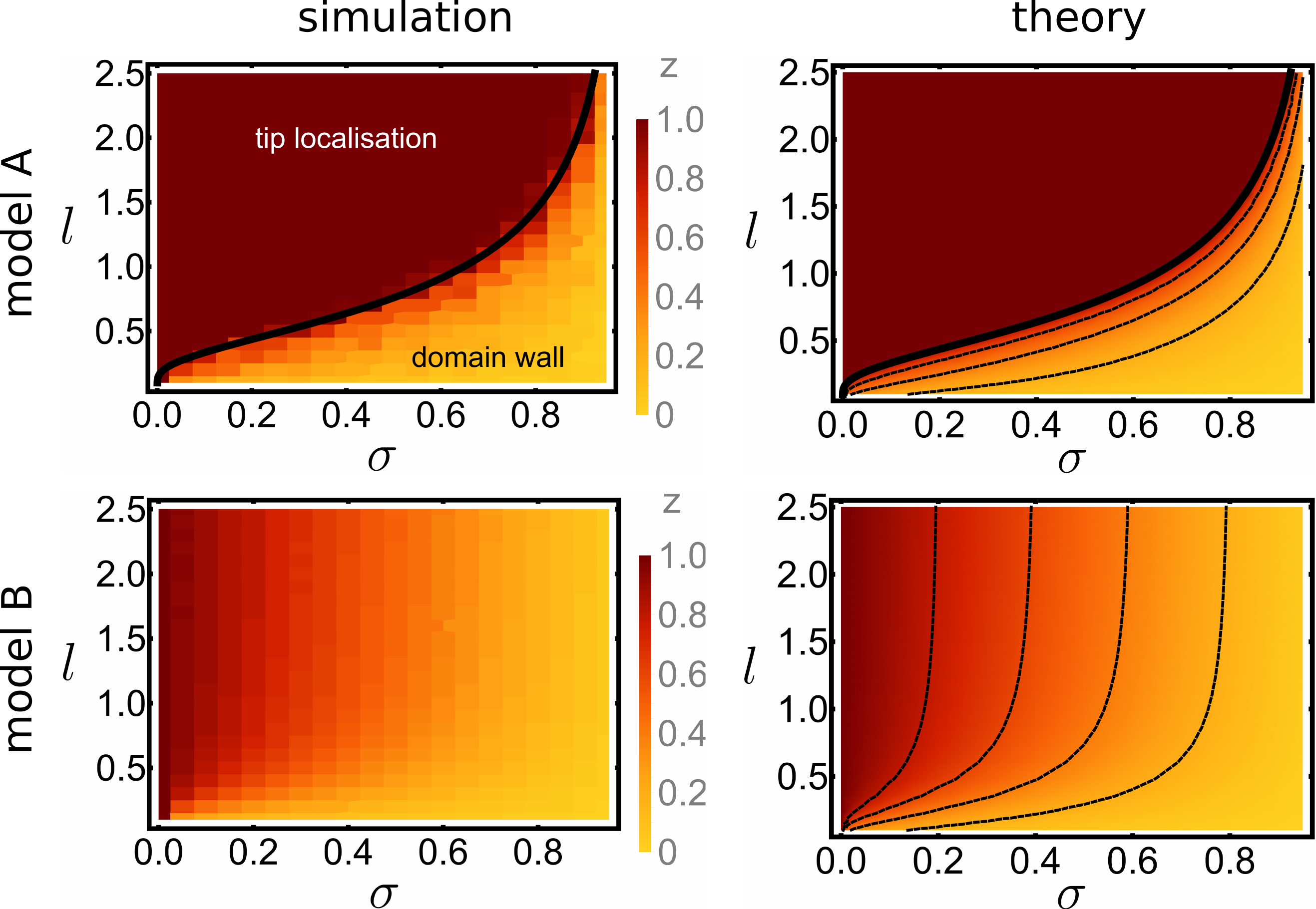}
    \caption{Phase diagrams for the DW position $z$ (colour-coded) for $L \,{=}\, 50$ and $\Omega \,{=}\, 0.001$ for model A (upper panel, $\beta \,{=}\, 0.2$) and model B (lower panel) as a function of the typical dwell length $l$ in the cytoplasm  and the cytoplasmic density at the protrusion base $\sigma$: the simulation results are shown on the left, the theoretical predictions for the DW position are shown on the right. The black thick lines in the diagrams for model A show the phase boundary $z \,{=}\, 1$ as obtained from theory. The dashed black lines in the diagrams obtained from theory are, from right to left respectively, contour lines of constant $z\,{=}\,0.2, 0.4, 0.6$ and additionally $z\,{=}\,0.8$ for model B.}
\label{fig:phasediagram}
\end{figure}

To address the first issue, we determine the average density  $\rho^{\mathrm{S}} (x)$ on the SSEP lane corresponding to the fully-localised-DW ansatz  by solving the bulk equation for the SSEP. For that, we  implement the continuum limit and, for model A, assume $\langle n^{\mathrm{T}}_{L} n^{\mathrm{S}}_{L} \rangle \,{\approx}\, \rho^{\mathrm{T}}_{L} \rho^{\mathrm{S}}_{L}$~\cite{Supplement}. Note, that we only need this MF approximation for the tip densities of model A. This is due to the otherwise symmetric attachment and detachment rates and the diffusive motion in the cytoplasm, for both of which the correlations drop out in the dynamical equation.  With the resulting equation for $\rho^{\mathrm{S}} (x)$ we can then conclude $z \,{=}\, 1 \,{-}\, l \cosh^{-1} \left[ \sigma \cosh (1/l) \right]$ and $z \,{=}\, 1 \,{-}\, l \sinh^{-1} \left[ \sigma \sinh (1/l) \right]$ for models A and B, respectively, where $l\,{:=}\, \sqrt{{D}/{\omega}}$ with $D\,{:=}\,\epsilon a^2$ being the diffusion constant in the cytoplasm~\cite{Supplement}; $l$ can be understood as the typical dwell length for motors in the cytoplasm before attaching to the filament. Comparing these expressions with our stochastic simulation results \cite{GILLESPIE1976403} we find excellent agreement [Fig.\ \ref{fig:phasediagram}]. 

The phase diagrams for the two models are qualitatively different. For model A, one can switch between the DW phase ($z \,{\ll}\, 1$) and the tip-localisation phase ($z \,{\approx}\, 1$) by only slightly increasing the dwell length $l$. In contrast, for model B, tip localisation is attained only as $\sigma \,{\rightarrow}\, 0$, even for large $l$~\cite{Supplement}.
This can be understood from the symmetry between the attachment and detachment processes in model B, which should also be reflected in a symmetry in occupancy between the filament and the cytoplasm. For large $l$, the cytoplasm becomes well-mixed with constant density $\sigma$. So, the average density on the filament is $\sigma$ as well, which is realised for a DW position $z \,{=}\, 1 \,{-}\, \sigma$. For model A, fast diffusion in the cytoplasm leads to rapid diffusion of the motors away from the tip. Therefore, the exit rate at the tip is high and motors quickly leave the tip of the filament, thus enabling tip localisation but reducing large jamming. For small but finite $\sigma$, there is tip localisation also if the typical dwell length is smaller than the system size, $l \,{\leq}\, 1$. That is, even if the motors have a relatively small $D$, tip localisation can occur if the tip has an enhanced detachment rate (model A). For model B, even by increasing $l$ well beyond the system size, tip localisation occurs only for very low motor density $\sigma$  at the base.

These features may have interesting biological implications: A higher detachment rate at the actin filament's end would promote tip localisation, but simultaneously avoid jamming. As a result, motor exchange between filament and cytoplasm occurs primarily near the tip (where the DW is located) and the motors at the tip can be continuously replenished by new ones delivered through the cytoplasm. Furthermore, energy consumption (ATP hydrolysis) is low, as the filament current is kept small and mainly restricted to the tip area. Transport to the tip is facilitated mainly by diffusion in the cytoplasm, which does not consume chemical energy. In summary, energy from ATP hydrolysis could efficiently be used to localise motors to the filament tip, while material transport is facilitated by diffusion in the cytoplasm~\cite{Supplement}. 

In this regard, the model is also interesting from a  theoretical point of view, as it allows for the calculation of the filament current $J^{\mathrm{T}}_i \,{=}\, \rho^{\text{T}}_i \,{-}\, f_i$ that depends on NN correlations or, equivalently, the NN covariances 
\begin{align}
\mathrm{Cov}_i := \langle n^{\mathrm{T}}_i n^{\mathrm{T}}_{i+1} \rangle - \rho^{\mathrm{T}}_i \rho^{\mathrm{T}}_{i+1} \,{=}\, f_i-\rho^{\mathrm{T}}_i \rho^{\mathrm{T}}_{i+1}
\end{align}
on the TASEP lane. To go beyond MF, we use  Eq.\ \eqref{fi}, which relates $f_i$ to the average densities, and find
$
\mathrm{Cov}_i \,{=}\, \rho^{\mathrm{T}}_i (1 \,{-}\, \rho^{\mathrm{T}}_{i+1}) \,{+}\, \omega \sum_{j=0}^i (\rho^{\mathrm{T}}_j \,{-}\, \rho^{\mathrm{S}}_j), 
$
which in the continuum limit translates to
\begin{align}
	\mathrm{Cov} (x) 
	= \rho^{\mathrm{T}} (x) \left[ 1{-}\rho^{\mathrm{T}} (x{+}a) \right] 
	+\Omega \int_{0}^x \! \mathrm{d}y  \, (\rho^{\mathrm{T}} {-} \rho^{\mathrm{S}}) 
\, .
\label{eq:cov_cont_limit}
\end{align} 
The value of the first term in Eq.~\eqref{eq:cov_cont_limit} depends sensitively on the width and shape of the DW (the density profile $\rho^{\mathrm{T}} (x)$ is increasing with $x$, so that $\rho^{\mathrm{T}} (x) (1 \,{-}\, \rho^{\mathrm{T}} (x{+}a))$ is maximal if both $\rho^{\mathrm{T}} (x)$ and $\rho^{\mathrm{T}} (x{+}a)$ are close to $0.5$). Therefore, one needs to refine the fully-localised-DW ansatz, by taking into account the stochastic dynamics of the DW position. Following Refs.~\cite{Derrida1995, 0305-4470-31-33-003} we consider the DW as a random walker with (site-dependent) hopping rates depending on currents and densities in the low- and high-density regions. For small $a$ we find
\begin{equation}
\rho^{\mathrm{T}} (x)  \approx \frac{\mathrm{erf} \left({(x-z)}/{W(z)} \right)+\mathrm{erf} \left({z}/{W(z)} \right)}{\mathrm{erf} \left({(1-z)}/{W(z)} \right)+\mathrm{erf} \left({z}/{W(z)} \right)} \, ,
\label{refined}
\end{equation}
where $W (z) \, {=} \,  \sqrt{2 \sigma a l \sinh \left(z/l \right)}$ ~\cite{Supplement}. This fluctuation-corrected DW profile, as well as the covariance obtained from it, agree very well with our simulation data [Fig.\ \ref{fig:DWcor}]. If one uses a refined MF method instead, accounting for second-order spatial derivatives, the DW width and the strength of the correlations are both markedly underestimated~\cite{Supplement}.

In general, covariances are non-zero only close to the DW, but there they can reach quite high values of around $0.2$. We have $\mathrm{Cov}_i \,{=}\,  \rho^{\mathrm{T}}_i (1 \,{-}\, \rho^{\mathrm{T}}_{i+1}) \,{-}\, J_i^\mathrm{T}$,  where the first term corresponding to the MF current dominates. Therefore,  the actual current $J^{\mathrm{T}}_i \,{=}\, \omega \sum_{j=0}^i (\rho^{\mathrm{S}}_j - \rho^{\mathrm{T}}_j)$ can be orders of magnitude smaller than the MF current, and scales with the particle exchange rate $\omega$~\cite{Supplement}. That is, density correlations dominate in such a way that they suppress the TASEP current substantially. This demonstrates that, even if basic properties of the single TASEP are captured well by MF theory, correlation effects in TASEP-based systems should be studied more closely and might lead to unanticipated features \cite{PhysRevE.67.066117, 1751-8121-41-9-095002, PhysRevLett.117.078102}. 

The covariances are non-negative everywhere, $\mathrm{Cov}_i \,{\geq}\, 0$, so the conditional probabilities obey the inequalities $\mathrm{Prob} \{ n^{\mathrm{T}}_i {=} 1 | n^{\mathrm{T}}_{i+1} {=} 1 \} \,{\geq}\, \mathrm{Prob} \{ n^{\mathrm{T}}_i {=} 1 \}$ and $\mathrm{Prob} \{ n^{\mathrm{T}}_{i+1} {=} 1 | n^{\mathrm{T}}_{i} {=} 1 \} \,{\geq}\, \mathrm{Prob} \{ n^{\mathrm{T}}_{i+1} {=} 1 \}$ at any site, implying that particles typically form clusters. As a result, from the DW region onwards (where the covariances are highest), the mean time a particle spends at a certain site is increased considerably compared to freely moving motors. This is due to the effective jump rate on the filament, which is decreased by excluded-volume effects. Thus, in case of tip localisation, particles spend more time near the tip than in the main part of the filament. This prolongation of the residence time is further enhanced by the exclusion in the cytoplasm that prevents motors from detaching if the cytoplasmic tip density is high. This is especially important for model A, where the cytoplasmic tip density takes a value of 1, which is much higher than in the bulk~\cite{Supplement}. Biologically, the extended residence time at the tip might facilitate the tasks of the motors or their cargo at the tip.

Our results all essentially rely on the exact moment-identity and the exact mass-balance equations. The derivation of both depends on the TASEP dynamics and the coupling between the two lanes, but not at all on the dynamics of the second (here SSEP) lane. Hence, those equations do not change if this dynamics on the second  lane is modified, and we expect them to be useful for other model systems in which a TASEP lane is coupled to another lattice via attachment and detachment kinetics. To our best knowledge, the moment-identity has not been mentioned before. We believe that it might open doors in the understanding of correlations and the prediction of active currents also in more elaborate models. Furthermore, both equations could easily be generalised to the case of different attachment and detachment rates. 

Our models could also be varied in other interesting ways. One could account for the three-dimensional geometry or for polymerisation and depolymerisation of the filament and the accompanying changes in length.
Nevertheless, we expect that some of the phenomena seen here should be robust against such modifications. Tip localisation, which is mostly based on mass conservation at the tip, should still be present. Second, the TASEP current might still be suppressed, and the roles of the TASEP and the diffusive lane in being responsible for tip localisation and motor transport, respectively, should remain untouched. This seems to be supported by more elaborate models in a related context~\cite{zhuravlev2, 1478-3975-11-1-016005}.

\begin{acknowledgments}
We thank Matthias Rank, Louis Reese, and Emanuel Reithmann for critical reading of this manuscript and for helpful discussions. 
This research was supported by the German Excellence Initiative via the program ``NanoSystems Initiative Munich'' (NIM) and by the Deutsche Forschungsgemeinschaft (DFG) through the Graduate School of Quantitative Biosciences Munich (QBM).
\end{acknowledgments}

\cleardoublepage
\pagebreak

\widetext
\begin{center}
\textbf{\large Supplemental Material: \\
Generic transport mechanisms for molecular traffic in cellular protrusions}
\end{center}
\setcounter{equation}{0}
\setcounter{figure}{0}
\setcounter{table}{0}
\setcounter{page}{1}
\makeatletter
\renewcommand{\theequation}{S\arabic{equation}}
\renewcommand{\thefigure}{S\arabic{figure}}
\renewcommand{\bibnumfmt}[1]{[S#1]}

This Supplemental Material gives details on the mathematical analysis of the lattice gas model. We will explain more thoroughly why the generic steady-state TASEP density profile is given by a domain wall and why there is no maximal current  phase. Furthermore, the calculation of the density profiles both for TASEP and SSEP are shown explicitly and the analytic expressions for the position of the domain wall are derived. It is demonstrated how domain wall theory is used concretely to improve on the mean-field TASEP density profiles, and on the prediction for the covariances. Those are relevant to see how the actual TASEP current differs from the expected mean-field current. We also display a comparison of the different currents (on the TASEP, on the SSEP and those for attachment and detachment) for a parameter set of model A where tip localisation occurs. Finally, we show that the steady-state behaviour of a geometry with several lanes for diffusion arranged on a cylinder around the TASEP lane can be reduced to the steady-state behaviour of our model by scaling the parameters for diffusion and attachment/detachment with the number of lanes for diffusion. The same holds true for a model where instead of exclusion on the lane for diffusion a finite carrying capacity $N_{\mathrm{max}} > 1$ is used.

\section{Calculation of the density profiles}
To set the stage, let us denote by $n^{\mu}_i$ the occupation number on site $i \in \{0, \ldots, L\}$ of lane $\mu \in \{\mathrm{T},\mathrm{S}\}$, i.e.\ we will write $n^{\mu}_i=0$ if site $i$ of lane $\mu$ is empty and $n^{\mu}_i=1$ if it is occupied ($\mathrm{T}$: TASEP, $\mathrm{S}$: SSEP). The common bulk master equations of the Markov processes, corresponding to model A and B as shown in Fig. \ref{fig:model}, are given by
\begin{align*}
\frac{\mathrm{d} \text{Prob} \{n^{\mathrm{T}}_i\}}{\mathrm{d} t}&= \text{Prob} \{n^{\mathrm{T}}_{i-1}, \overline{n^{\mathrm{T}}_{i}}\} - \text{Prob} \{n^{\mathrm{T}}_{i}, \overline{n^{\mathrm{T}}_{i+1}} \} 
+ \omega \left(\text{Prob} \{n^{\mathrm{S}}_{i}, \overline{n^{\mathrm{T}}_{i}}\} - \text{Prob} \{n^{\mathrm{T}}_{i}, \overline{n^{\mathrm{S}}_i}\}\right),\\
\frac{\mathrm{d} \text{Prob} \{n^{\mathrm{S}}_i\}}{\mathrm{d} t} &= \epsilon \left(\text{Prob} \{n^{\mathrm{S}}_{i-1}, \overline{n^{\mathrm{S}}_i}\} + \text{Prob} \{n^{\mathrm{S}}_{i+1}, \overline{n^{\mathrm{S}}_{i}}\} -  \text{Prob} \{n^{\mathrm{S}}_{i}, \overline{n^{\mathrm{S}}_{i+1}}\}-\text{Prob} \{n^{\mathrm{S}}_{i}, \overline{n^{\mathrm{S}}_{i-1}}\}\right) +\\
&+ \omega \left(\text{Prob} \{n^{\mathrm{T}}_{i}, \overline{n^{\mathrm{S}}_{i}}\} - \text{Prob} \{n^{\mathrm{S}}_{i}, \overline{n^{\mathrm{T}}_{i}}\}\right),
\end{align*}
where $\text{Prob}\{n^{\mu}_{i}\}$  denotes the probability that $n^{\mu}_{i}=1$ and $\text{Prob}\{n^{\mu}_{i}, \overline{n^{\nu}_j}\}$ the one that $n^{\mu}_{i}=1$ and $n^{\nu}_{j}=0$. The term $\text{Prob} \{n^{\mathrm{T}}_{i-1}, \overline{n^{\mathrm{T}}_{i}}\} - \text{Prob} \{n^{\mathrm{T}}_{i}, \overline{n^{\mathrm{T}}_{i+1}} \} $  is due to the jump process on the TASEP lane that respects the exclusion property and occurs at bare rate $\nu=1$. The terms proportional to $\omega$ describe the exchange between the lanes, again respecting the exclusion. And finally, the term proportional to $\epsilon$ describes the diffusion on the SSEP lane. 
Note that we assume exclusion not only on the filament but also in the cytoplasm. This is based on the idea that, due to the finite size of particles, there should be a maximal number inside any finite volume element, introducing a carrying capacity $N_{\mathrm{max}}$. If we assume that the \textit{maximal} effective attachment and detachment rate stay the same, i.e.\ if we assume that attachment happens at rate $\omega \left( n^{\mathrm{S}} \left( 1-n^{\mathrm{T}} \right) \right)$  and detachment at rate $\omega \left( n^{\mathrm{T}} \left( N_{\mathrm{max}} -n^{\mathrm{S}} \right) \right)$ where $n^{\mathrm{S}} \in \{0, 1, \dots, N_{\mathrm{max}}\}$, the case $N_{\mathrm{max}}$ finite but arbitrary can be reduced to  $N_{\mathrm{max}}=1$ by redefinition of the parameters, and in the following we will focus only on the case $N_{\mathrm{max}}=1$. We will, however, come back to this case again in the last paragraph of the Supplemental Material when we discuss the case of several lanes for diffusion.

The bulk master equations can be rewritten in terms of averages over the occupation numbers by using that $\langle n^{\mu}_i \rangle = \mathrm{Prob} \left(n^{\mu}_i =1 \right)=\mathrm{Prob} \{n^{\mu}_i \}$ and  $\langle n^{\mu}_i \left( 1- n^{\nu}_i \right) \rangle = \mathrm{Prob} \{n^{\mu}_i, \overline{n^{\nu}_i} \}$ hold, as each site can be either empty or occupied by one particle/motor:
\begin{align}
&\partial_t \rho^{\mathrm{T}}_i= \rho^{\mathrm{T}}_{i-1} - f_{i-1} - \rho^{\mathrm{T}}_{i} + f_{i}  
+ \omega \left(\rho^{\mathrm{S}}_{i} - \rho^{\mathrm{T}}_{i}\right), \label{TASEPdynamics}\\
&\partial_t \rho^{\mathrm{S}}_i= \epsilon \left(\rho^{\mathrm{S}}_{i+1}+\rho^{\mathrm{S}}_{i-1}-2 \rho^{\mathrm{S}}_{i} \right)
+ \omega \left(\rho^{\mathrm{T}}_{i} - \rho^{\mathrm{S}}_{i}\right), \label{SSEPdynamics}
\end{align}
where $f_i=\left\langle n^{\mathrm{T}}_i n^{\mathrm{T}}_{i+1} \right\rangle$ is the nearest-neighbour correlator for the TASEP.  Summing both equations we find
\begin{align}
\partial_t (\rho^{\mathrm{T}}_i+\rho^\mathrm{S}_i)= \rho^{\mathrm{T}}_{i-1} - f_{i-1} - \rho^{\mathrm{T}}_{i} + f_{i} + \epsilon \left(\rho^{\mathrm{S}}_{i+1}+\rho^{\mathrm{S}}_{i-1}-2 \rho^{\mathrm{S}}_{i} \right) \label{combineddensity}
\end{align}
for the time evolution of the combined density $\rho^{\mathrm{T}}_i+\rho^{\mathrm{S}}_i$ in the bulk. At the left boundary (base) we find
\begin{align}
&\partial_t \rho^{\mathrm{T}}_0= - \rho^{\mathrm{T}}_{0} + f_{0}  
+ \omega \left(\rho^{\mathrm{S}}_{0} - \rho^{\mathrm{T}}_{0}\right) \label{base} \\
&\partial_t \rho^{\mathrm{S}}_0= \alpha (1-\rho^{\mathrm{S}}_0)-\epsilon \rho^{\mathrm{S}}_0 +\epsilon \left(\rho^{\mathrm{S}}_{1}- \rho^{\mathrm{S}}_{0} \right)
+ \omega \left(\rho^{\mathrm{T}}_{0} - \rho^{\mathrm{S}}_{0}\right) \nonumber
\end{align}
for both models. For the TASEP lane there is only outflux from site $0$ to site $1$ or exchange with site $0$ of the SSEP lane. On the SSEP lane, there is influx at rate $\alpha$ (respecting the exclusion at site 0 of the SSEP), outflux with the diffusion rate $\epsilon$, diffusion between site $0$ and site $1$ and exchange with site 0 of the TASEP. The behaviour at the right boundary (tip) differs between the models and is given by
\begin{align}
&\partial_t \rho^{\mathrm{T}}_L= \rho^{\mathrm{T}}_{L-1} - f_{L-1}  
- \beta \left(\rho^{\mathrm{T}}_{L} - \left\langle n^{\mathrm{T}}_L n^{\mathrm{S}}_L \right\rangle \right) \label{tipmodelA}\\
&\partial_t \rho^{\mathrm{S}}_L= \epsilon \left( \rho^{\mathrm{S}}_{L-1}-\rho^{\mathrm{S}}_{L}\right) + \beta \left(\rho^{\mathrm{T}}_{L} - \left\langle n^{\mathrm{T}}_L n^{\mathrm{S}}_L \right\rangle \right) \nonumber
\end{align}
for model A, and by
\begin{align}
&\partial_t \rho^{\mathrm{T}}_L= \rho^{\mathrm{T}}_{L-1} - f_{L-1}  
+ \omega \left(\rho^{\mathrm{S}}_{L} - \rho^{\mathrm{T}}_{L}\right) \label{tipmodelB}\\
&\partial_t \rho^{\mathrm{S}}_L= \epsilon \left( \rho^{\mathrm{S}}_{L-1}-\rho^{\mathrm{S}}_{L}\right)
+ \omega \left(\rho^{\mathrm{T}}_{L} - \rho^{\mathrm{S}}_{L}\right)  \nonumber
\end{align}
for model B. For model A, at site $L$ of the TASEP lane there is influx from the neighbouring site $L-1$ and outflux to site $L$ of the SSEP lane  at bare rate $\beta$ respecting the exclusion.  For site $L$ of the SSEP lane there is diffusion between sites $L-1$ and $L$ of the SSEP lane and influx from site $L$ of the TASEP lane. For model B, we have the same behaviour except that the asymmetric exchange between sites $L$ of the TASEP and SSEP lane is replaced by the symmetric exchange $\omega \left(\rho^{\mathrm{T}}_{L} - \rho^{\mathrm{S}}_{L}\right) $. Since the exchange terms drop out when considering the time derivative of $\rho^{\mathrm{T}}_L+\rho^{\mathrm{S}}_L$, at the tip of both models it holds
\begin{align}
&\partial_t (\rho^{\mathrm{T}}_L+\rho^{\mathrm{S}}_L)= \rho^{\mathrm{T}}_{L-1} - f_{L-1}  + \epsilon \left( \rho^{\mathrm{S}}_{L-1}-\rho^{\mathrm{S}}_{L}\right)= J^{\mathrm{T}}_{L-1}+J^{\mathrm{S}}_{L-1},
\label{combineddensityL}
\end{align}
and at the base
\begin{align}
&\partial_t (\rho^{\mathrm{T}}_0+\rho^{\mathrm{S}}_0)= - \rho^{\mathrm{T}}_{0} + f_{0}  + \alpha (1-\rho^{\mathrm{S}}_0)-\epsilon \rho^{\mathrm{S}}_0 +\epsilon \left(\rho^{\mathrm{S}}_{1}- \rho^{\mathrm{S}}_{0} \right)=-J^{\mathrm{T}}_0+  \alpha (1-\rho^{\mathrm{S}}_0)-\epsilon \rho^{\mathrm{S}}_0 -J^{\mathrm{S}}_0
\label{combineddensity0}
\end{align}
where we introduced the local currents 
\begin{align}
J^{\mathrm{T}}_i=\rho^{\mathrm{T}}_{i} - f_{i} \label{TASEPcurrent}
\end{align}
and
\begin{align} 
J^{\mathrm{S}}_i=\epsilon \left( \rho^{\mathrm{S}}_{i}-\rho^{\mathrm{S}}_{i+1}\right)\label{SSEPcurrent}
\end{align}
for $i=0, \dots, L-1$ on the filament and in the cytoplasm, respectively.  With that Eq. (\ref{combineddensity}) translates to
\begin{align}
\partial_t (\rho^{\mathrm{T}}_i+\rho^\mathrm{S}_i)=J^{\mathrm{T}}_{i-1}+J^{\mathrm{S}}_{i-1}-J^{\mathrm{T}}_{i}-J^{\mathrm{S}}_{i}.
\label{combineddensitybulk}
\end{align}
Note that we can introduce currents $J^{\mathrm{T/S}}_{-1}$ and $J^{\mathrm{T/S}}_{L}$ as well. However, as the system is closed at the tip $J^{\mathrm{T/S}}_{L}\equiv 0$, and since there is no direct influx from the left into the TASEP $J^{\mathrm{T}}_{-1}=0$. In order to be consistent with the structure of Eq. (\ref{combineddensitybulk}) where the currents appear as $J_{i-1}-J_i$, we define 
\begin{align}
J^{\mathrm{S}}_{-1}=\alpha (1-\rho^{\mathrm{S}}_0)-\epsilon \rho^{\mathrm{S}}_0
\end{align}
so that
\begin{align*}
&\partial_t (\rho^{\mathrm{T}}_0+\rho^{\mathrm{S}}_0)=-J^{\mathrm{T}}_0+  J^{\mathrm{S}}_{-1} -J^{\mathrm{S}}_0
\end{align*}
holds.

As mentioned above, we are interested in the steady-state behaviour of the system and therefore set $\partial_t \rho^{T/S}_i \equiv 0$ for all $i$. In particular, we have
\begin{align*}
0=\partial_t \sum_{i=0}^L (\rho^{\mathrm{T}}_i+\rho^{\mathrm{S}}_i)=J^{\mathrm{S}}_{-1}
\end{align*}
where we used Eqs. (\ref{combineddensityL}), (\ref{combineddensity0}), (\ref{combineddensitybulk}). As a result,
\begin{align}
\rho^{\mathrm{S}}_0 = \frac{\alpha}{\alpha+\epsilon} := \sigma,
\label{SSEPboundary0}
\end{align}
and the parameter $\sigma$ as used for the phase diagrams in Fig. \ref{fig:phasediagram} can be identified as the motor density in the cell body.

To proceed, let us now use the continuum limit where $a=\frac{1}{L}\rightarrow 0$. That is, we will replace the discrete lattice by the continuous space $[0,1]$ and the average occupation numbers are replaced by a continuous density
\begin{align*}
\rho^{\mu}_i \rightarrow \rho^{\mu} \left(x=x_i\right) \hspace{10pt} \mathrm{where} \hspace{10pt} x_i= i a.
\end{align*}
With that, we can substitute $\rho^{\mu} (x_i) \pm a \partial_x \rho^{\mu} (x_i) +\frac{1}{2} a^2 \partial_x^2 \rho^{\mu} (x_i) + \mathcal{O} (a^3)$ for $\rho^{\mu}_{i\pm1}$ and the currents are
\begin{align*}
J^{\mathrm{T}} (x) &= \rho^{\mathrm{T}} (x) - f(x) \\
J^{\mathrm{S}} (x) &= - a \epsilon \partial_x \rho^{\mathrm{S}} (x) - \frac{1}{2} a^2 \epsilon \partial_x^2 \rho^{\mathrm{S}} (x) + \mathcal{O} (a^3)
\end{align*}
where $f(x)$ is the continuous version of $f_i$. In the steady-state Eq. (\ref{combineddensitybulk}) translates to
\begin{align*}
0= - a \partial_x (J^{\mathrm{T}}+J^{\mathrm{S}}) + \frac{1}{2} a^2 \partial_x^2 (J^{\mathrm{T}}+J^{\mathrm{S}})  + \mathcal{O} (a^3) = \\
=-a \partial_x (\rho^{\mathrm{T}} (x) - f(x)) + \mathcal{O} (a^2)
\end{align*}
to first order in $a$ and thus, we have $\rho^{\mathrm{T}} (x) - f(x) = const$ to lowest order in $a$. Using Eq. (\ref{combineddensityL}) or (\ref{combineddensity0}) we conclude that $const =0$ or that the combined current $J^{\mathrm{T}}_i + J^{\mathrm{S}}_i$ must be constant and zero everywhere. This corresponds to the fact that the system is closed at the tip so that in steady state on average there is no influx into the system from the base. Therefore,
\begin{align*}
\rho^{\mathrm{T}} (x) \equiv f(x)
\end{align*}
or, equivalently,
\begin{align*}
\text{Prob} \{n^{\mathrm{T}}_i\} = \rho^{\mathrm{T}}_i = f_i = \left\langle n^{\mathrm{T}}_i n^{\mathrm{T}}_{i+1} \right\rangle = \text{Prob} \{ n^{\mathrm{T}}_i, n^{\mathrm{T}}_{i+1} \} = \text{Prob} \{ n^{\mathrm{T}}_{i+1} | n^{\mathrm{T}}_{i} \} \text{Prob} \{ n^{\mathrm{T}}_i\}
\end{align*}
holds to lowest order  in $a$ where we defined $\text{Prob} \{ n^{\mathrm{T}}_i, n^{\mathrm{T}}_{i+1} \}=\text{Prob} \left( n^{\mathrm{T}}_i=1, n^{\mathrm{T}}_{i+1} =1\right)$ and $\text{Prob} \{ n^{\mathrm{T}}_{i+1} | n^{\mathrm{T}}_{i} \} =\text{Prob} \left( n^{\mathrm{T}}_{i+1}=1 | n^{\mathrm{T}}_{i}=1 \right) $ analogously to before. This implies that either 
\begin{align}
\rho^{\mathrm{T}}_i=\text{Prob} \{n^{\mathrm{T}}_i\} = 0
\label{zerodensity}
\end{align}
or
\begin{align*} 
\text{Prob} \{ n^{\mathrm{T}}_{i+1} | n^{\mathrm{T}}_{i} \}=1.
\end{align*}
The latter implies that, whenever site $i$ is occupied, site $i+1$ is occupied with probability 1 as well. Hence, if there is some site $j$ occupied at some time, any site $i>j$ is occupied at that time as well. It follows that to lowest order in $a$ the density suddenly jumps from zero (\ref{zerodensity}) to maximal density
\begin{align}
\rho^{\mathrm{T}}_i=1.
\label{maximaldensity}
\end{align}
This means that to lowest order the steady-state TASEP profile is given by a step function separating a region of $0$ density on the left (towards the base) from a region of density $1$ on the right (towards the tip) and, in particular, there is no maximal current phase as it occurs for TASEP alone. Instead, the generic TASEP profile is given by a domain wall and we make the following fully-localised-DW ansatz:
\begin{align}
\rho^{\mathrm{T}} (x) =
	\begin{cases}
	0 \hspace{10pt} &\text{for} \hspace{10pt} x \in {[0,z}[ \\
	1 \hspace{10pt} &\text{for} \hspace{10pt} x \in {]z,1]}
	\end{cases} 
	\label{ansatzrho1supplement}
\end{align}
where $z\in[0,1]$ is the position of the domain wall (step) that we will determine later. 

We will now use this ansatz to determine the steady-state density profile for the SSEP lane depending on $z$. For this purpose, let us go back to Eq. (\ref{SSEPdynamics}) that is given by
\begin{align*}
0=\epsilon a^2 \partial_x^2 \rho^{\mathrm{S}} (x) + \Omega a (\rho^{\mathrm{T}} (x) - \rho^{\mathrm{S}} (x))
\end{align*}
in the continuum limit in the steady state. This equation can be solved in the two regions $x<z$ and $x>z$ as
\begin{align*}
\rho^{\mathrm{S}} (x) =
	\begin{cases}
	A_1 \cosh \left(\sqrt{\frac{\Omega}{\epsilon a}} x\right) +A_2 \sinh  \left(\sqrt{\frac{\Omega}{\epsilon a}} x\right) \hspace{10pt} &\text{for} \hspace{10pt} x \in {[0,z}[ \\
	1+B_1 \cosh  \left(\sqrt{\frac{\Omega}{\epsilon a}} x \right) +B_2 \sinh \left (\sqrt{\frac{\Omega}{\epsilon a}} x \right) \hspace{10pt} &\text{for} \hspace{10pt} x \in {]z,1]}.
	\end{cases} 
\end{align*}
The constants $A_1$, $A_2$, $B_1$ and $B_2$ can be determined from the boundary conditions: Eq. (\ref{tipmodelA}) yields $\beta  \left(\rho^{\mathrm{T}}_{L} - \left\langle n^{\mathrm{T}}_L n^{\mathrm{S}}_L \right\rangle \right) =0$ to lowest order in $a$ for model A, and assuming that $n^{\mathrm{T}}_L$ and $n^{\mathrm{S}}_L$ are uncorrelated (mean-field assumption that is supported by the simulations) we conclude that
\begin{align*}
\rho^{\mathrm{S}}_L=1
\end{align*}
for model A, unless $\rho^{\mathrm{T}}_{L}=0$ and so $z=1$.
For model B, we can use Eq. (\ref{tipmodelB}) and find
\begin{align*}
\partial_x \rho^{\mathrm{S}} (1) =\frac{\Omega}{\epsilon} \left(1-\rho^{\mathrm{S}} (1) \right),
\end{align*}
again unless $z=1$. This latter case needs to be treated separately. This can be done for instance by regarding the "domain wall`` as a boundary layer with slope $(\rho^{\mathrm{S}} (1)_{\mathrm{right}} - \rho^{\mathrm{S}} (1)_{\mathrm{left}})/a$ where $\rho^{\mathrm{S}} (1)_{\mathrm{right}}$ is the density at the very last site $L$ and $\rho^{\mathrm{S}} (1)_{\mathrm{left}}$ is the limit $x \rightarrow 1$ of the low-density phase. Basically, this case can be understood as the limit where the domain wall is shifted to the right out of the system. As long as $\epsilon \in \mathcal{O}(a^0)$, the generic density profile is given by a domain wall separating density $0$ on the left from density $1$ on the right, that is $\rho^{\mathrm{T}} \in \{0,1\}$ holds to lowest order in $a$. In the limiting cases $\rho^{\mathrm{T}}\equiv 1$ the density reaches the value of $1$ in only a few lattice sites from the base, whereas in the case $\rho^{\mathrm{T}}\equiv 0$ the density has a very small spike (boundary layer) only at the tip. The details of this calculation are, however, out of the scope of this letter and we will carry on with the treatment of the case $z\neq 1$.

\begin{figure}[t]
    \centering
    \includegraphics[trim={0cm 0cm 0cm 0cm},clip,height=4.8cm]{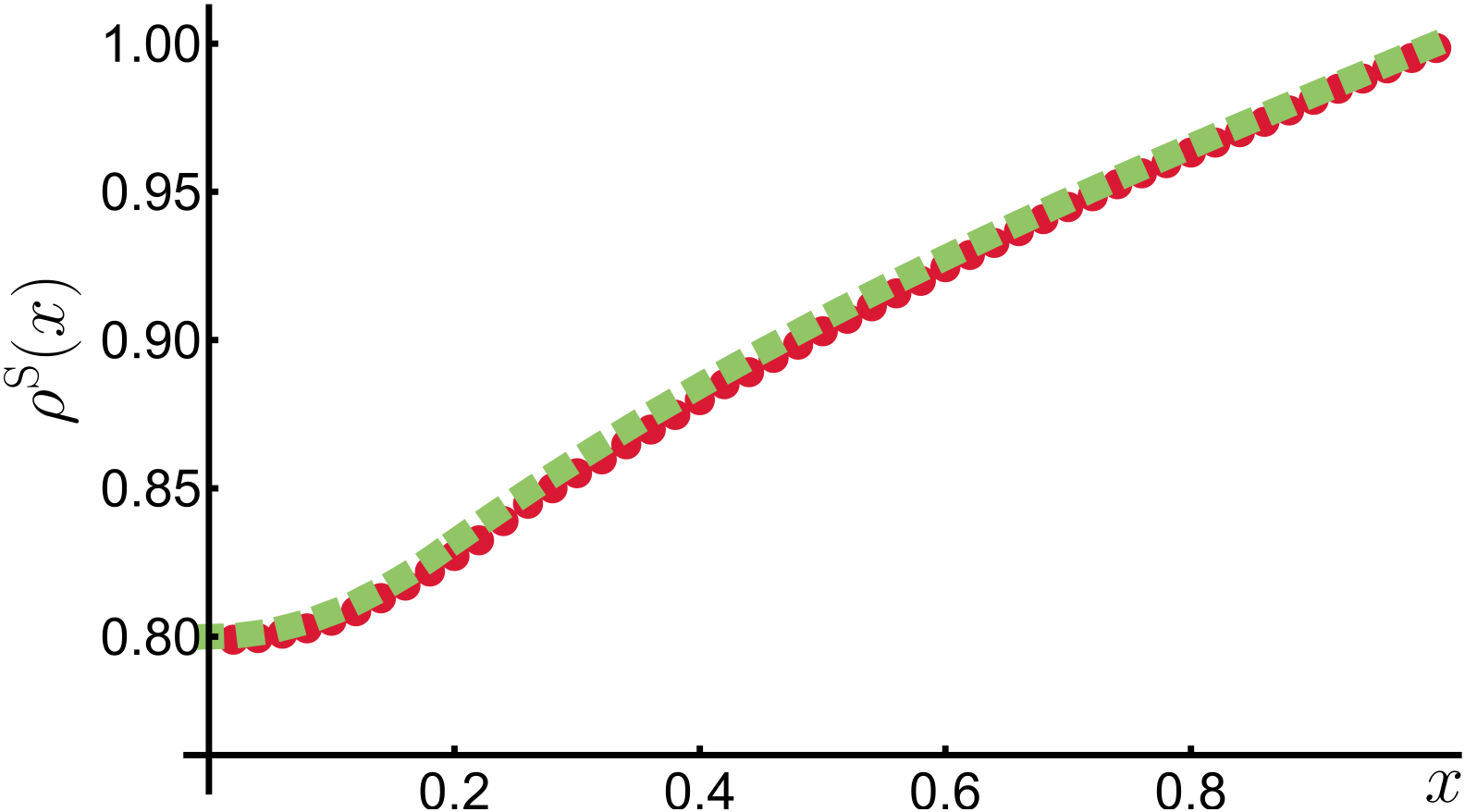}\hfill
    \includegraphics[trim={0cm 0cm 0cm 0cm},clip,height=4.8cm]{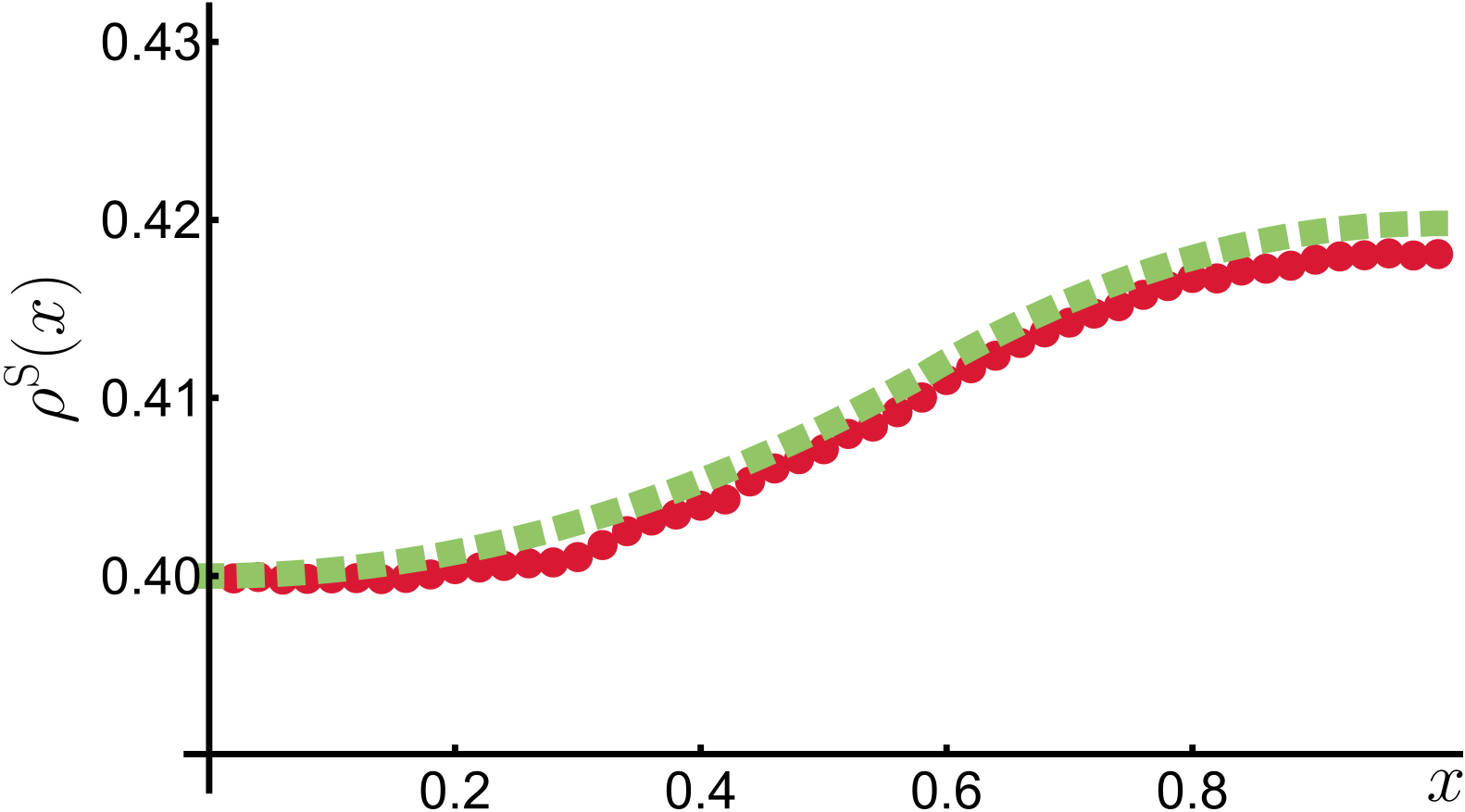}
    \caption{Steady-state SSEP density profiles are shown examplarily for model A (left panel) with $L \,{=}\, 50$, $\beta \,{=}\, 0.2$, $\Omega \,{=}\, 0.001$, $\alpha \,{=}\, 0.1$ and $\epsilon \,{=}\, 0.025$ and for model B (right panel) with $L \,{=}\, 50$, $\Omega \,{=}\, 0.001$, $\alpha \,{=}\, 0.2$ and $\epsilon \,{=}\, 0.3$. The simulation results (filled red circles) agree well with the theoretical prediction (dotted green curve) according to Eq. \eqref{SSEPdensityA} and \eqref{SSEPdensityB}, respectively. For model A, the density is increasing towards the tip where it has a high slope and reaches a value of 1. For model B, the density is more homogeneous and has a small slope at the tip.}
    \label{fig:SSEPprofiles}
\end{figure}

For both models we have Eq. \eqref{SSEPboundary0} for the boundary condition at the base. Requiring that both the diffusive steady-state density profile and the derivative thereof is continuous (it can be seen from Eq. (\ref{SSEPdynamics}) that $ \left(\rho^{\mathrm{S}}_{i+1}-\rho^{\mathrm{S}}_{i} \right) - \left(\rho^{\mathrm{S}}_{i}-\rho^{\mathrm{S}}_{i-1} \right)  = \mathcal{O} (a)$ holds), the following expression for the SSEP profiles can be derived:
\begin{align}
\rho^{\mathrm{S}} (x) =
	\begin{cases}
	\sigma \cosh \left( \frac{x}{l}\right) + \left( \coth \left( \frac{1}{l} \right) \left( \cosh \left( \frac{z}{l} \right) - \sigma \right) - \sinh \left( \frac{z}{l}\right) \right) \sinh  \left(\frac{x}{l}\right) \hspace{10pt} &\text{for} \hspace{10pt} x \in {[0,z}[  \\
	1-\left( \cosh \left( \frac{z}{l}\right) - \sigma \right) \left( \cosh \left( \frac{x}{l} \right) - \coth \left( \frac{1}{l} \right) \sinh \left(  \frac{x}{l} \right) \right) \hspace{10pt} &\text{for} \hspace{10pt} x \in {]z,1]}
	\end{cases}
	\label{SSEPdensityA}
\end{align}
for model A and
\begin{align}
\rho^{\mathrm{S}} (x) =
	\begin{cases}
	\sigma \cosh \left(\frac{x}{l}\right) + \left( \gamma \left( \cosh \left( \frac{z}{l} \right) - \sigma \right) - \sinh \left( \frac{z}{l}\right) \right) \sinh  \left( \frac{x}{l}\right) \hspace{10pt} &\text{for} \hspace{10pt} x \in {[0,z}[   \\
	1-\left( \cosh \left( \frac{z}{l}\right) - \sigma \right) \left( \cosh \left( \frac{x}{l} \right) - \gamma \sinh \left( \frac{x}{l} \right) \right) \hspace{10pt} &\text{for} \hspace{10pt} x \in {]z,1]}
	\end{cases}
	\label{SSEPdensityB}
\end{align}
for model B where 
\begin{align*}
l=\sqrt{\frac{\epsilon a}{\Omega}}=\sqrt{\frac{\epsilon a^2}{\omega}}=\sqrt{\frac{D}{\omega}}
\end{align*}
with the diffusion constant $D=\epsilon a^2$ and
\begin{align*}
\gamma= \frac{\sinh \left( \frac{1}{l}\right) +\frac{a}{l} \cosh \left(\frac{1}{l} \right)}{ \cosh \left( \frac{1}{l} \right) +\frac{a}{l} \sinh \left(\frac{1}{l}\right)} \approx \tanh \left( \frac{1}{l} \right).
\end{align*}
To illustrate the differences between the two models, Fig. \ref{fig:SSEPprofiles} shows typical density profiles on the SSEP lane for both cases. As one can see, the density profiles differ significantly between the two models. For model A, the density generically increases considerably from the protrusion base with density $\sigma$ towards the tip with density 1 (unless $z=1$). In contrast, the density for model B at the tip reaches a value that is only slightly larger than the value at the base. The fact that for model A the cytoplasmic density at the tip reaches a value very close to 1 is also the reason why for model A, the exclusion in the cytoplasm has a much higher influence on the residence time at the tip compared to model B where the cytoplasm is occupied quite homogeneously and where exclusion at the tip is not more important than in the bulk.

Using Eqs. \eqref{SSEPdensityA} and \eqref{SSEPdensityB} we now have expressions for the steady-state density profiles that only depend on one quantity, namely $z$, the position of the domain wall. Using the mass-balance equations [Eq. \eqref{model1conservation} and the corresponding one for model B], this enables us to find an analytic formula for $z$.

\section{Position of the domain wall}

In order to determine the position of the domain wall, we go back to Eq. (\ref{TASEPdynamics}). In the steady state this reduces to
\begin{align*}
f_i -f_{i-1}=\rho^{\mathrm{T}}_i-\rho^{\mathrm{T}}_{i-1} + \omega \left( \rho^{\mathrm{T}}_i - \rho^{\mathrm{S}}_i \right).
\end{align*}
Hence, we can write
\begin{align*}
f_i = f_0 + \sum_{j=1}^i \left( f_j-f_{j-1}\right)=f_0 + \sum_{j=1}^i \left( \rho^{\mathrm{T}}_j-\rho^{\mathrm{T}}_{j-1} + \omega \left( \rho^{\mathrm{T}}_j - \rho^{\mathrm{S}}_j \right) \right) = f_0 + \rho^{\mathrm{T}}_i-\rho^{\mathrm{T}}_0 + \omega \sum_{j=1}^i \left( \rho^{\mathrm{T}}_j - \rho^{\mathrm{S}}_j \right).
\end{align*}
Using Eq. (\ref{base}) we can rewrite $f_0=\rho^{\mathrm{T}}+\omega \left( \rho^{\mathrm{T}}_0 - \rho^{\mathrm{S}}_0 \right)$ and, thus, 
\begin{align}
f_i =  \rho^{\mathrm{T}}_i + \omega \sum_{j=0}^i \left( \rho^{\mathrm{T}}_j - \rho^{\mathrm{S}}_j \right). \label{fisupplement}
\end{align}
This is the moment-identity [Eq. (\ref{fi})]. It is an important result for us as it gives an exact relation between the nearest-neighbour correlator $f_i = \left\langle n^{\mathrm{T}}_i n^{\mathrm{T}}_{i+1} \right\rangle$ and the average densities. We will use it later to predict the covariances and the TASEP current in the system. This will allow us to make a sharp distinction between the actual TASEP current $J^{\mathrm{T}}_i$ and the mean-field current $J^{\mathrm{T}}_{\mathrm{MF},i} = \rho^{\mathrm{T}}_i \left( 1 - \rho^{\mathrm{T}}_{i+1}\right)$.

From the moment-identity, Eq. \eqref{fisupplement}, we can also easily deduce the mass-balance equations [Eq. (\ref{model1conservation}) and the corresponding one for model B] by using the boundary conditions at the tip [Eq. (\ref{tipmodelA}) for model A and (\ref{tipmodelB}) for model B] and the moment-identity for $i=L-1$:
\begin{align*}
\mathrm{model \ A:} \hspace{20pt} &\omega \sum_{i=0}^{L-1} \left( \rho^{\mathrm{T}}_i - \rho^{\mathrm{S}}_i \right)= - \beta \left( \rho^{\mathrm{T}}_L - \left\langle n^{\mathrm{T}}_L n^{\mathrm{S}}_L \right\rangle \right) \\
\mathrm{model \ B:} \hspace{20pt} &\omega \sum_{i=0}^L \left( \rho^{\mathrm{T}}_i - \rho^{\mathrm{S}}_i \right)=0.
\end{align*} 

In order to determine $z$ we will now insert the domain wall ansatz, Eq. \eqref{ansatzrho1supplement}, for the TASEP together with the respective SSEP profile  into the respective mass-balance equation. For this purpose, we need to write the mass-balance equations in the continuum limit: Using that $\int_0^1 \mathrm{d} x \ \rho^{\mathrm{T}} (x) = 1-z$ and that $\sum_{i=0}^L \rho^{\mu}_i \approx \int_0^L \mathrm{d} i \ \rho^{\mu} (i a) = \frac{1}{a} \int_0^1 \mathrm{d} x \ \rho^{\mu} (x)$ we find to lowest order in $a$:
\begin{align*}
\mathrm{model \ A:} \hspace{20pt} &\Omega \left(1-z-\int_0^1 \mathrm{d} x \ \rho^{\mathrm{S}} (x) \right) = - \beta \left( \rho^{\mathrm{T}} (1) \left( 1 -  \rho^{\mathrm{S}} (1) \right) \right) =-\epsilon a \partial_x \rho^{\mathrm{S}} (1)\\
\mathrm{model \ B:} \hspace{20pt} &  1-z - \int_0^1 \mathrm{d}x \ \rho^{\mathrm{S}} (x) =0
\end{align*} 
where for model A the mean-field assumption  $\left\langle n^{\mathrm{T}}_L n^{\mathrm{S}}_L \right\rangle \approx \rho^{\mathrm{T}}_L \rho^{\mathrm{S}}_L$ was used again, together with Eq. (\ref{tipmodelA}).
Inserting the explicit densities (\ref{SSEPdensityA}) or (\ref{SSEPdensityB}) into these expressions, integrating and cancelling common factors results in
\begin{align}
\mathrm{model \ A:} \hspace{20pt} &\sigma \cosh \left( \frac{1}{l} \right) = \cosh \left( \frac{ 1-z}{l} \right) \label{positionofdwAsupplement}\\
\mathrm{model \ B:} \hspace{20pt} &\sigma \sinh \left( \frac{1}{l} \right) = \sinh \left( \frac{1-z}{l} \right).\label{positionofdwBsupplement}
\end{align}
Those directly lead to the expressions for the domain wall given in the main text:
\begin{align}
\mathrm{model \ A:} \hspace{20pt} &z=1-l \cosh^{-1} \left( \sigma \cosh \left( \frac{1}{l} \right) \right) \label{positionofdwAsupplementexplicit}\\
\mathrm{model \ B:} \hspace{20pt} &z=1-l \sinh^{-1} \left( \sigma \sinh \left( \frac{1}{l} \right) \right).\label{positionofdwBsupplementexplicit}
\end{align}
We see that $z$ is shifted towards the tip (base) for smaller (higher) values of $\sigma := \frac{\alpha}{\alpha+\epsilon}$. It is interesting to note that for model A the limit $z\nearrow 1$ is reached for finite values of the parameters $\sigma$, $\omega$ and $\epsilon$ when $\sigma \cosh \left( \frac{1}{l} \right) =1$ or $l=1/\cosh^{-1} (1/\sigma)$, whereas for model B, $z \nearrow 1$ is only possible in the limit $\sigma \rightarrow 0$. This is also reflected in the fact that $\sigma \cosh \left( \frac{1}{l} \right) = \cosh \left( \frac{ 1-z}{l} \right)$ lacks a real solution for $z$ if $\sigma \cosh \left( \frac{1}{l} \right) <1$. That is, for model A there are parameter regimes where there is no domain wall and where to lowest order the density is zero everywhere. For model B, this case only occurs in the limit where the cytoplasmic density at the base is zero. Even for large $l$ we have $z=1-l \sinh^{-1} \left( \sigma \sinh \left( \frac{1}{l} \right) \right) \approx 1- l \sinh^{-1} \left( \frac{\sigma}{l}  \right) \approx 1-\sigma$ and thus, $z<1$ unless $\sigma=0$.


\section{Refined mean-field TASEP density profile}
Certainly, the TASEP profile is not just given by a plain step function but has a smooth form that separates the two density regions via an intermediate region where the density increases strongly but whose width is finite. This is due to fluctuations in the stochastic system that soften this transition.
Our first approach to resolve this finite width relies on an idea similar to the method used in \cite{saha}, namely to split the system into a part around the domain wall and the parts further away from it. By this, only parts of the terms contribute, respectively, and assuming a steep domain wall for the TASEP one can neglect the terms stemming from the exchange with the diffusive lane in this narrow region: Reconsidering equation (\ref{TASEPdynamics}) we realize that in the steady state  it can be approximated as
\begin{align*}
0=a \partial_x \left( \rho^{\mathrm{T}} \left( \rho^{\mathrm{T}} -1\right)\right) + \frac{1}{2} a^2 \partial_x^2 \rho^{\mathrm{T}} + \Omega a \left( \rho^{\mathrm{S}} - \rho^{\mathrm{T}} \right)
\end{align*}
where we used the continuum limit and a mean-field assumption for the nearest-neighbours on the TASEP: $f_i =\left\langle n^{\mathrm{T}}_i n^{\mathrm{T}}_{i+1} \right\rangle \approx \rho^{\mathrm{T}}_i \rho^{\mathrm{T}}_{i+1}$.
If we want to investigate the vicinity of the domain wall $x \approx z$ this can be done by considering $\tilde{x}:= (x-z)/a$ that is very large away from the domain wall. Using this coordinate system the above equation looks like follows:
\begin{align*}
0= \partial_{\tilde{x}} \left( \rho^{\mathrm{T}} (\tilde{x}) \left( \rho^{\mathrm{T}} (\tilde{x}) -1\right)\right) + \frac{1}{2} \partial_{\tilde{x}}^2 \rho^{\mathrm{T}} (\tilde{x}) + \Omega a \left( \rho^{\mathrm{S}} (\tilde{x})- \rho^{\mathrm{T}} (\tilde{x}) \right).
\end{align*}
Dropping the term of order $\mathcal{O} (a)$, we end up with
\begin{align*}
\rho^{\mathrm{T}} (\tilde{x}) \left( \rho^{\mathrm{T}} (\tilde{x}) -1\right) + \frac{1}{2} \partial_{\tilde{x}} \rho^{\mathrm{T}} (\tilde{x}) = const
\end{align*}
where $const$ can be estimated from the boundary conditions for $\tilde{x}\rightarrow \pm \infty$: we have $\rho^{\mathrm{T}} (x=0)=1-\rho^{\mathrm{T}} (x=1) =0$ and $\left. \partial_x \rho^{\mathrm{T}} (x) \right|_{x=0/1} \approx 0$, and hence, using $\tilde{x}\rightarrow \pm \infty$ we find that $const \approx 0$. As a result, after integrating $\rho^{\mathrm{T}} (\tilde{x}) \left( \rho^{\mathrm{T}} (\tilde{x}) -1\right) + \frac{1}{2} \partial_{\tilde{x}} \rho^{\mathrm{T}} (\tilde{x}) =0$ with respect to $\tilde{x}$, using that $\rho^{\mathrm{T}} (z) =\frac{1}{2}$, and rescaling we find
\begin{align*}
\rho^{\mathrm{T}} (x) \approx \frac{1}{2} \left[ 1+\tanh \left( \frac{x-z}{a}\right) \right]
\end{align*}
for the refined mean-field TASEP density profile. Contrary to the step profile, Eq. \eqref{ansatzrho1supplement}, it exhibits a finite width that scales like $a$ or inversely proportionally to $L$. But as one can infer from Fig. \ref{fig:DWcor} it still underestimates the actual width due to fluctuations that are ignored by the mean-field assumption for the nearest-neighbours on the TASEP.
\section{Domain wall theory}

This is why we also pursue another approach to refine our prediction for the domain wall, namely by going beyond mean-field theory and treating the domain wall as a random walker that moves in a non-uniform potential with reflecting boundaries.
This technique has been introduced in Ref. \cite{Derrida1995} for the TASEP and, more generally, for shock waves in Ref. \cite{0305-4470-31-33-003}. It has been applied to TASEP-LK in  Refs. \cite{PhysRevLett.90.086601, PhysRevE.70.046101} using non-uniform hopping rates. 
We follow their ideas and describe the domain wall by site-dependent hopping rates
\begin{align}
w_{l,i} = \frac{J_{\mathrm{LD},i}}{\rho^{\mathrm{T}}_{\mathrm{HD},i}-\rho^{\mathrm{T}}_{\mathrm{LD},i}} \hspace{40pt} w_{r,i} = \frac{J_{\mathrm{HD},i}}{\rho^{\mathrm{T}}_{\mathrm{HD},i}-\rho^{\mathrm{T}}_{\mathrm{LD},i}} \label{hoppingrates}
\end{align}
to the left and right from site $i$, respectively. Here, $J_{\mathrm{LD},i}$ ($J_{\mathrm{HD},i}$) denotes the steady-state TASEP current at site $i$ under the assumption that site $i$ is in the low-density (high-density) phase, i.e.\ on the left (right) of the domain wall. $\rho^{\mathrm{T}}_{\mathrm{LD},i}$ and $\rho^{\mathrm{T}}_{\mathrm{HD},i}$ are the respective TASEP densities in the low- and high-density phase. The heuristic reason why the hopping rates have the above form, Eq. \eqref{hoppingrates}, is that the low-density current $J_{\mathrm{LD},i}$ should just correspond to the current arriving at the left of the domain wall and causing the domain wall to move one step to the left. Thereby, at site $i$ the density increases by $\rho^{\mathrm{T}}_{\mathrm{HD},i}-\rho^{\mathrm{T}}_{\mathrm{LD},i}$ at rate $w_{l,i}$. Similarly, the high-density current $J_{\mathrm{HD},i}$ should correspond to the current leaving the region at the right of the domain wall, and causing the domain wall to move one step to the left. This happens at rate $w_{r,i}$ and the density decreases by $\rho^{\mathrm{T}}_{\mathrm{HD},i}-\rho^{\mathrm{T}}_{\mathrm{LD},i}$. 

From the detailed balance condition $p_{s,i} w_{r,i}=p_{s,i+1} w_{l,i+1}$ it is easy to see that the stationary distribution of the shock position $p_{s,i}$ satisfies
\begin{align}
p_{s,i} \propto \frac{1}{w_{l,i}} \exp \left[-\sum_{j=1}^{i-1} \ln \left( \frac{w_{l,j}}{w_{r,j}}\right) \right].\label{shockdistribution}
\end{align}
We will show next that $w_{l,i_z}=w_{r,i_z}$
holds for $i_z$ being the discrete domain wall position ($i_z=z/a$): From Eq. (\ref{TASEPdynamics}) as well as from the definition of the local TASEP current, Eq. \eqref{TASEPcurrent}, we know that $J^{\mathrm{T}}_i = J^{\mathrm{T}}_{i-1} + \omega \left( \rho^{\mathrm{S}}_i - \rho^{\mathrm{T}}_i \right)$. Hence, we have
\begin{align*}
J^{\mathrm{T}}_i &= J^{\mathrm{T}}_0 +\omega \sum_{j=1}^i \left(\rho^{\mathrm{S}}_j - \rho^{\mathrm{T}}_j \right), \hspace{20pt} \mathrm{or}\\
J^{\mathrm{T}}_i &= J^{\mathrm{T}}_{L-1} -\omega \sum_{j=i+1}^{L-1} \left(\rho^{\mathrm{S}}_j - \rho^{\mathrm{T}}_j \right).
\end{align*}
As mentioned after introducing $J_{\mathrm{LD/HD},i}$ in Eq. \eqref{hoppingrates}, $J_{\mathrm{LD/HD},i}$ denotes the TASEP current at site $i$ assuming that this site is in the low-density (high-density) region. That is, we can calculate $J_{\mathrm{LD},i}$  ($J_{\mathrm{HD},i}$) by using the above iterations starting from $i=0$ ($i=L-1$) and assuming that between site $0$ (site $L-1$) and site $i$ the densities are given in the low-density (high-density) phase:
\begin{align}
J_{\mathrm{LD},i} &= J_{\mathrm{LD},0} +\omega \sum_{j=1}^i \left(\rho^{\mathrm{S}}_{\mathrm{LD},j} - \rho^{\mathrm{T}}_{\mathrm{LD},j} \right), \hspace{20pt} \mathrm{or} \label{currentLD}\\
J_{\mathrm{HD},i} &= J_{\mathrm{HD},L-1} -\omega \sum_{j=i+1}^{L-1} \left(\rho^{\mathrm{S}}_{\mathrm{HD},j} - \rho^{\mathrm{T}}_{\mathrm{HD},j} \right).\label{currentHD}
\end{align}
We want to compare those two currents right at the domain wall $i=i_z$. For this purpose, we use that  $J_{\mathrm{LD},j}=J^{\mathrm{T}}_j$ for $j<i_z$ and $J_{\mathrm{HD},j}=J^{\mathrm{T}}_j$ for $j>i_z$ since we chose $z$ in such a way that the densities to the left (right) of the fixed domain wall $i_z$ are in the low-density (high-density) phase. As a result,
\begin{align*}
J_{\mathrm{HD},i_z}-J_{\mathrm{LD},i_z}= J_{\mathrm{HD},L-1} -\omega \sum_{j=i_z+1}^{L-1} \left(\rho^{\mathrm{S}}_{\mathrm{HD},j} - \rho^{\mathrm{T}}_{\mathrm{HD},j} \right) -J^{\mathrm{T}}_0 -\omega \sum_{j=1}^{i_z} \left(\rho^{\mathrm{S}}_j - \rho^{\mathrm{T}}_j \right)=J^{\mathrm{T}}_{L-1} - \omega \sum_{j=0}^{L-1} \left(\rho^{\mathrm{S}}_j - \rho^{\mathrm{T}}_j \right)=0
\end{align*}
for both models:  $J^{\mathrm{T}}_{L-1} = \beta \left( \rho^{\mathrm{T}}_L - \left\langle n^{\mathrm{T}}_L n^{\mathrm{S}}_L \right\rangle \right)$ for model A and $J^{\mathrm{T}}_{L-1}=\omega \left( \rho^{\mathrm{T}}_L -\rho^{\mathrm{S}}_L \right)$ for model B and, hence, the above holds due to the respective mass-balance equation. It follows that 
\begin{align}
w_{l,i_z} = w_{r,i_z}\label{currentequality}
\end{align}
and, thus, $\ln \left( \frac{w_{l,i_z}}{w_{r,i_z}}\right)=0$. Similarly, we have $\ln \left( \frac{w_{l,j}}{w_{r,j}}\right)<0$ for $j<i_z$ and $\ln \left( \frac{w_{l,j}}{w_{r,j}}\right)>0$ for $j>i_z$ such that the domain wall preferably walks towards the (fixed) domain wall position $i_z$ than away from it. Therefore, the exponential in Eq. (\ref{shockdistribution}) has a maximum at $i=i_z$ and taking the continuous version of this equation
\begin{align}
p_s (x) \propto \frac{1}{w_l (x)} \mathrm{exp} \left[-\frac{1}{a} \int_0^x \mathrm{d} x' \ln \left( \frac{w_l (x')}{w_r (x')}\right) \right]
\label{shockdistributioncontinuous}
\end{align}
we can use the method of steepest descent to approximate
\begin{align*}
\frac{1}{w_l (x)} \mathrm{exp} \left[-\frac{1}{a} \int_0^x \mathrm{d} x' \ln \left( \frac{w_l (x')}{w_r (x')}\right)\right] \approx \frac{1}{w_l (z)} \mathrm{exp} \left[-\frac{1}{a} \int_0^z \mathrm{d} x' \ln \left( \frac{w_l (x')}{w_r (x')}\right) \right] \mathrm{exp} \left[-\frac{1}{2 a} (x-z)^2 \partial_x \left. \ln \left( \frac{w_l (x)}{w_r (x)}\right) \right|_{x=z} \right]
\end{align*}
where $w_l (x)$ and $w_r (x)$ is the continuous version of $w_{l,i}$ and $w_{r,i}$, respectively. Since $\frac{w_l (x)}{w_r (x)}=\frac{J_{\mathrm{LD}} (x)}{J_{\mathrm{HD}} (x)}$, we conclude
\begin{align}
p_s (x) &\propto \mathrm{exp} \left[-\frac{1}{2 a} (x-z)^2 \frac{J_{\mathrm{LD}}' (z) J_{\mathrm{HD}} (z) - J_{\mathrm{HD}}' (z) J_{\mathrm{LD}} (z)}{J_{\mathrm{LD}} (z) J_{\mathrm{HD}} (z)}\right]= \mathrm{exp} \left[-\frac{1}{2 a} (x-z)^2 \frac{J_{\mathrm{LD}}' (z) - J_{\mathrm{HD}}' (z)}{J_{\mathrm{LD}} (z) }\right]= \label{shockdistributionfinal}\\
&=\mathrm{exp} \left[-\frac{(x-z)^2}{W(z)^2} \right] \nonumber
\end{align}
where we used that $J_{\mathrm{LD}} (z)=J_{\mathrm{HD}} (z)$ [Eq. (\ref{currentequality})] and defined the width $W(z)$ by  
\begin{align*}
W^{-2} (z) = \frac{J_{\mathrm{LD}}' (z)- J_{\mathrm{HD}}' (z)}{2 a J_{\mathrm{LD}} (z)}. 
\end{align*}
To find a formula for $\rho^{\mathrm{T}} (x)$ we realize that 
\begin{align*}
\rho^{\mathrm{T}} (x)=\rho^{\mathrm{T}}_{\mathrm{HD}} (x) \int_0^x \mathrm{d} \tilde{x} \ p_s (\tilde{x}) +\rho^{\mathrm{T}}_{\mathrm{LD}} (x) \int_x^1 \mathrm{d} \tilde{x} \ p_s (\tilde{x}) \approx \int_0^x \mathrm{d} \tilde{x} \ p_s (\tilde{x})
\end{align*}
since as along as the shock position is left (right) of $x$ there is a high-density (low-density) region at $x$ and since we can approximate $\rho^{\mathrm{T}}_{\mathrm{HD}} (x) \approx 1$ and $\rho^{\mathrm{T}}_{\mathrm{LD}} (x) \approx 0$ to lowest order. Using Eq. (\ref{shockdistributionfinal}) we can determine $\rho^{\mathrm{T}} (x)$ as
\begin{align}
\rho^{\mathrm{T}} (x) \approx \frac{\int_0^x \mathrm{d} \tilde{x} \ \mathrm{exp} \left(-\frac{(\tilde{x}-z)^2}{W(z)^2} \right) }{\int_0^1 \mathrm{d} \tilde{x} \ \mathrm{e} \left(-\frac{(\tilde{x}-z)^2}{W(z)^2} \right)} = 
\frac{\mathrm{erf} \left( \frac{x-z}{W(z)}\right)+\mathrm{erf} \left( \frac{z}{W(z)}\right)}{\mathrm{erf} \left( \frac{1-z}{W(z)}\right)+\mathrm{erf} \left( \frac{z}{W(z)}\right)} \label{resultdomainwallprofile}
\end{align}
where $\mathrm{erf}$ is the error function. Here, we used that we need to normalize $p_s (x)$ such that $\int_0^1 \mathrm{d} \tilde{x} \ p_s (\tilde{x}) =1$.

In order to use this formula we still need to determine the width $W(z)$ more concretely. For this purpose, we must make some assumption on how we treat the diffusive (SSEP) lane when considering the low- and high-density currents. We are not aware of previous attempts that apply domain wall theory in case where a TASEP is coupled to another lattice that is occupied stochastically as well. Then the attachment and detachment rates for the TASEP not only depend on the TASEP occupancy but also on the occupancy on the other lattice. The difficulty with this situation is that it is a priori not clear if one should assume that the occupancy on the coupled lattice is fixed, that is, does not change very much if the domain wall is shifted towards the left or right, or if one should look at the momentary steady-state density corresponding to a certain position of the domain wall. For simplicity, we chose the first ansatz that seems to be working well. 
That means, we use Eqs. (\ref{SSEPdensityA}) and (\ref{SSEPdensityB}) for the SSEP density profile with the calculated fixed position of the domain wall $z$ given by Eqs. (\ref{positionofdwAsupplement}) and (\ref{positionofdwBsupplement}) and we assume that this SSEP density does not change considerably when the domain wall is shifted shortly to some other position $\neq z$, i.e.\ $\rho^{\mathrm{S}}_{\mathrm{LD}} (x) = \rho^{\mathrm{S}}_{\mathrm{HD}} (x) = \rho^{\mathrm{S}} (x)$.

With that in mind, using Eqs. (\ref{currentLD}) and (\ref{currentHD}), we find that
\begin{align*}
J_{\mathrm{LD}} (x) = \Omega \int_0^x \mathrm{d} \tilde{x} \ \rho^{\mathrm{S}} (\tilde{x})
\end{align*}
and obtain
\begin{align*}
\mathrm{model \ A:} \hspace{20pt} &J_{\mathrm{HD}} (x) =- \Omega \int_x^1 \mathrm{d} \tilde{x} \ \left(\rho^{\mathrm{S}} (\tilde{x})-1\right)-\epsilon a \partial_x \rho^{\mathrm{S}} (1)\\
\mathrm{model \ B:} \hspace{20pt} &J_{\mathrm{HD}} (x) =- \Omega \int_x^1 \mathrm{d} \tilde{x} \ \left(\rho^{\mathrm{S}}  (\tilde{x})  -1\right)
\end{align*}
where we used that $\rho^{\mathrm{T}}_{\mathrm{LD}} (x) \approx 0$ and $\rho^{\mathrm{T}}_{\mathrm{HD}} (x)\approx 1$.
As a result, $J_{\mathrm{LD}}' (x) = \Omega \rho^{\mathrm{S}} (x)$ and $J_{\mathrm{HD}}' (x) = \Omega \left( \rho^{\mathrm{S}} (x)-1\right)$ so that
\begin{align*}
W (z) = \sqrt{2 a  \int_0^z \mathrm{d} \tilde{x} \ \rho^{\mathrm{S}} (\tilde{x})}=\sqrt{2 \sigma a l \sinh \left( \frac{z}{l} \right)}
\end{align*}
where the last equality results from integrating $\rho^{\mathrm{S}}$ and then using the defining equation for $z$. Interestingly, the result in this form agrees for both models.

We observe that the width $W(z)$ increases with increasing $z$ or decreasing distance from the tip. Furthermore, in the limit $a \rightarrow 0$ it holds that $l \rightarrow 0$ (if we keep $\epsilon$ constant) so that we can approximate $\mathrm{arccosh} \left( \sigma \cosh \left( \frac{1}{l} \right) \right) \approx \frac{1}{l} + \ln \left( \sigma \right)$ and, similarly, $\mathrm{arcsinh} \left( \sigma \sinh \left( \frac{1}{l} \right) \right) \approx \frac{1}{l} + \ln \left( \sigma \right)$. Thus, $z \approx -l \ln \left( \sigma \right)$ for both models in the limit $a\rightarrow 0$ [Eqs. \eqref{positionofdwAsupplementexplicit}, \eqref{positionofdwBsupplementexplicit}]. With that we find
\begin{align*}
W (z) \approx \frac{a^{\frac{3}{4}} \epsilon^{\frac{1}{4}}}{\Omega^{\frac{1}{4}}} \left(1-\sigma^2  \right)^{\frac{1}{2}}
\end{align*}
in the limit $a\rightarrow 0$ and so the width of the domain wall decreases with decreasing $a$ or increasing number of sites $L$.

Note that in order to calculate the position of the domain wall $z$ by the mass-balance equation, we can use the step function instead of the refined profiles. This is due to the fact that there the errors more or less cancel since only the sum $\sum_{i=0}^L \rho^{\mathrm{T}}_i$ over the densities on the left and on the right of the domain wall enters. However, this approximation gets worse the closer the calculated $z$ is to 0 or 1 since then, the errors are not symmetric anymore. As a result, also the approximation, Eq. \eqref{resultdomainwallprofile}, deteriorates. Certainly, if there is no solution for $z$, e.g.\ for some parameters in model A, we can not use this ansatz either. Furthermore, if $\epsilon$ is too large, we expect our approximation that $\rho^{\mathrm{S}} (x)$ does not change when the domain wall is shifted to deteriorate as well.


\section{Covariances and Currents}

\begin{figure}[t]
    \centering
    \includegraphics[trim={0cm 0cm 0cm 0cm},clip,height=4.8cm]{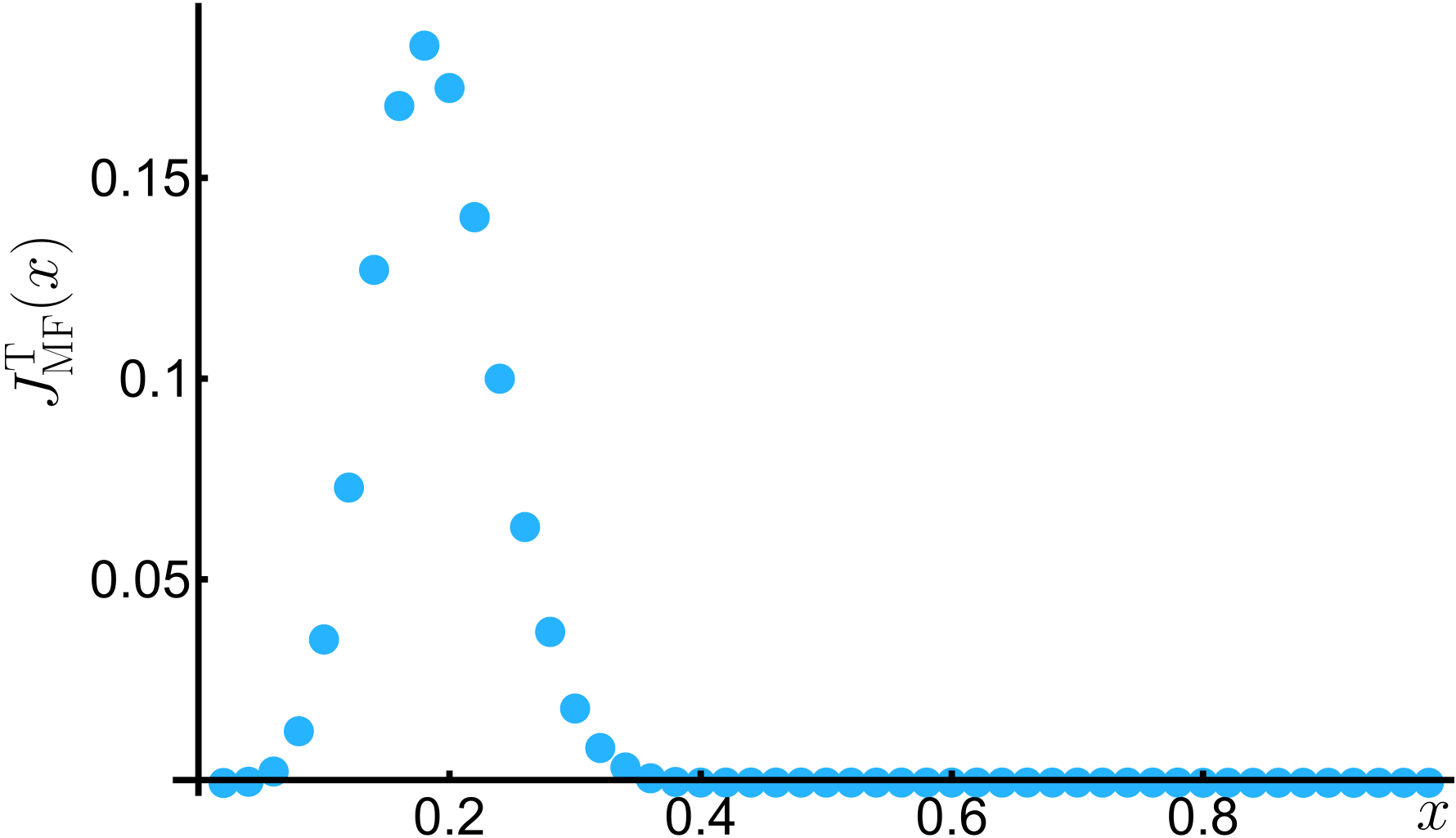} \hfill
    \includegraphics[trim={0cm 0cm 0cm 0cm},clip,height=4.8cm]{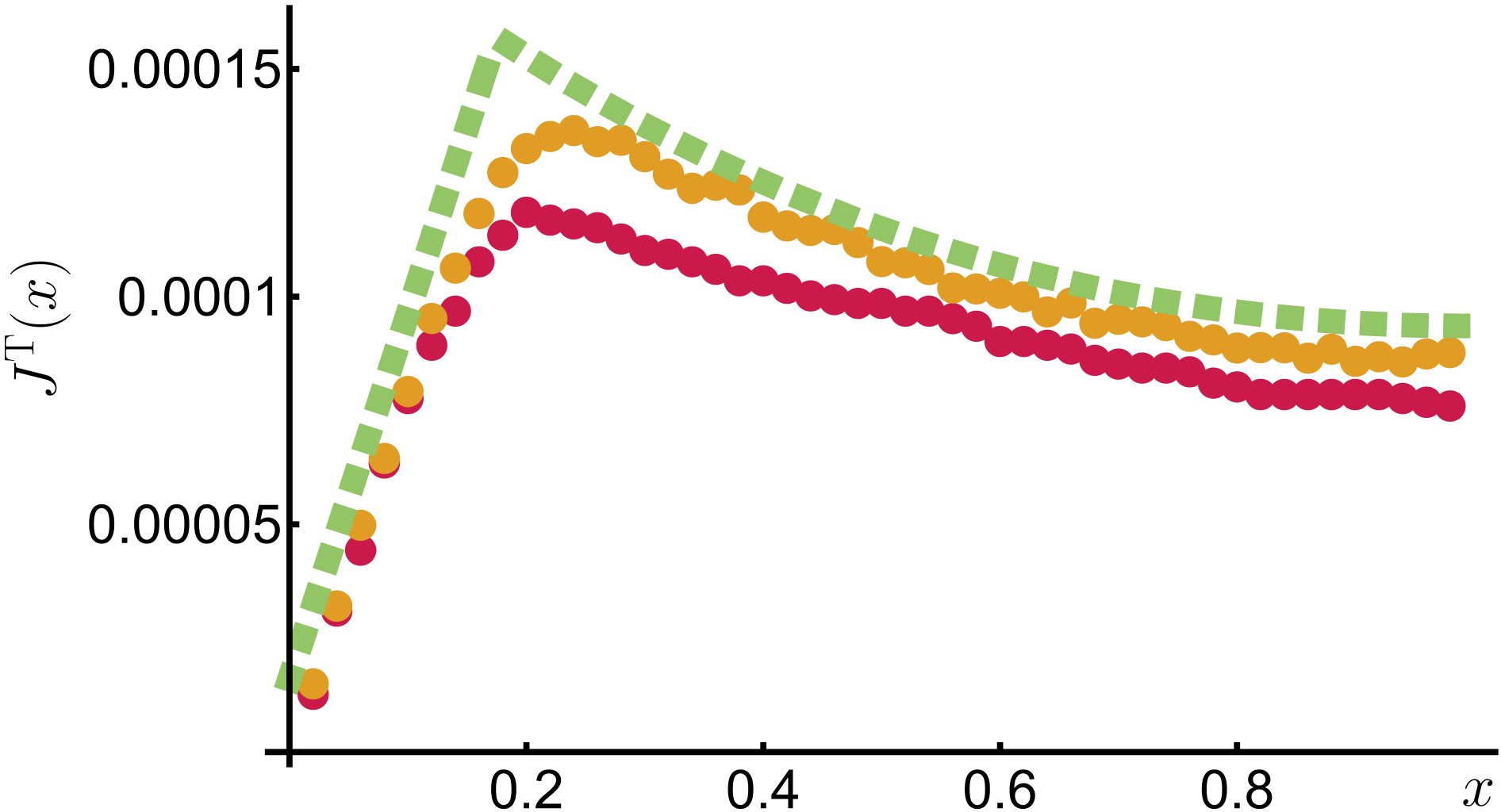}
    \caption{Comparison between the mean-field current (left panel) and the actual current (right panel) for  model A with $L=50$, $\beta=0.2$, $\Omega=0.001$, $\alpha=0.1$ and $\epsilon=0.025$. The mean-field current (filled blue (grey) circles) is obtained from the average densities in a simulation. On the right, we show the actual currents as measured directly in a simulation (filled red (dark grey) circles), as obtained from the average densities and the nearest neighbour correlations in the simulation using $J^{\mathrm{T}}_i = \rho^{\mathrm{T}}_i - f_i$ (filled orange (grey) circles) and as predicted by the theory from Eq. (\ref{actualcurrent}) (dotted green (light grey) curve).}
    \label{fig:comparecurrentsupplement}
\end{figure}

\begin{figure}[t]
    \centering
    \includegraphics[trim={0cm 0cm 0cm 0cm},clip,height=4.8cm]{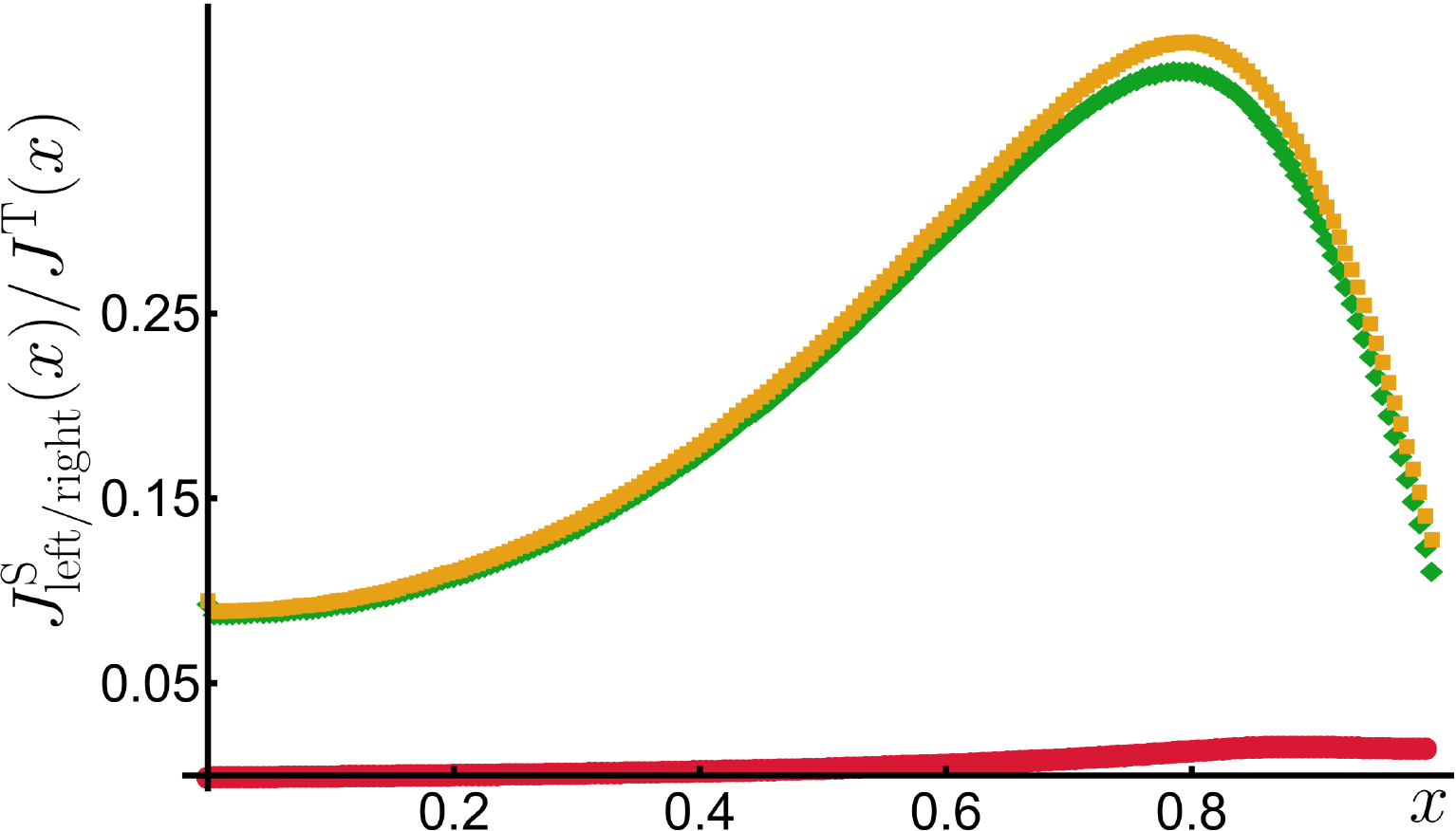} \hfill
    \includegraphics[trim={0cm 0cm 0cm 0cm},clip,height=4.8cm]{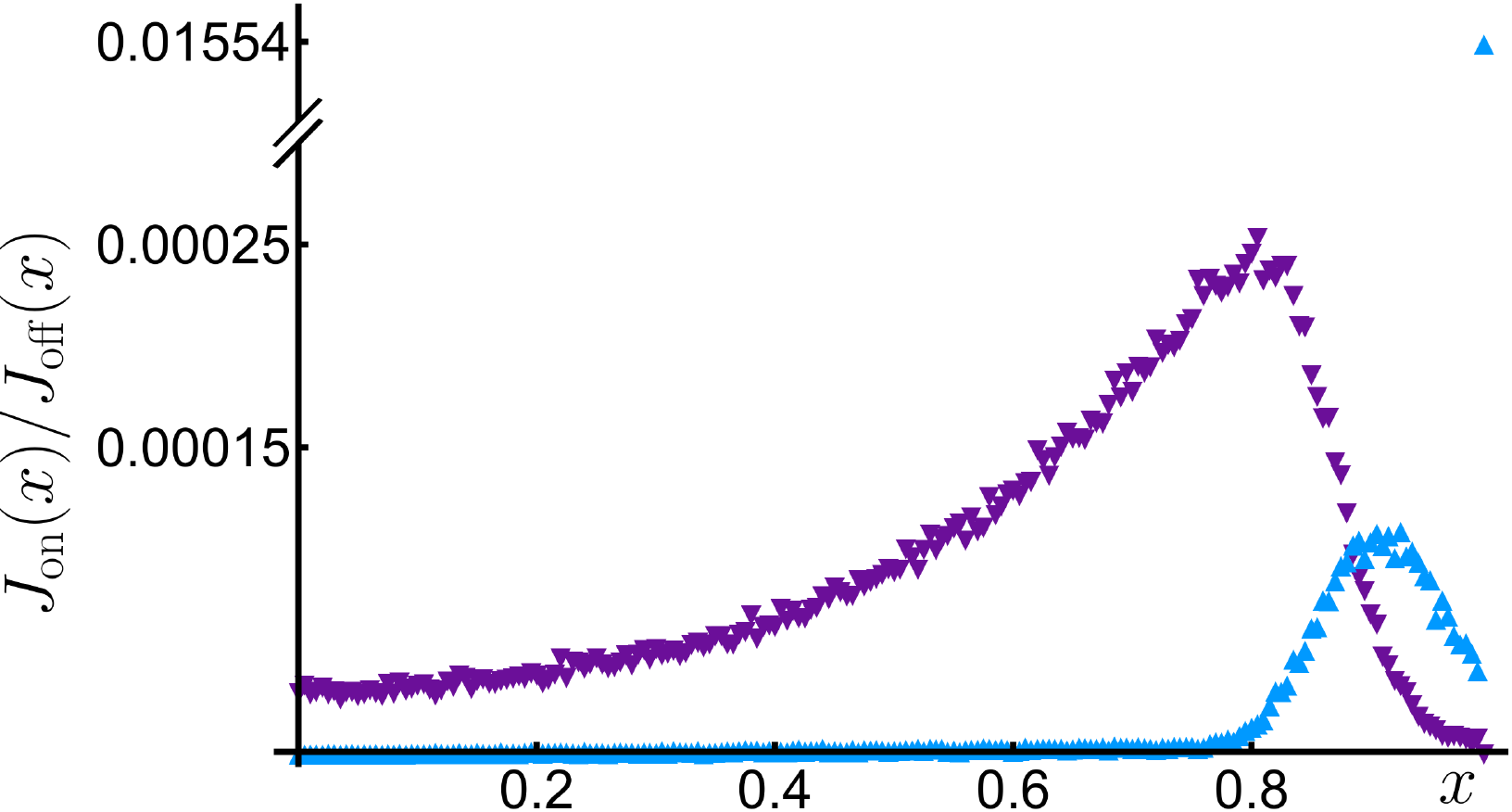}
    \caption{Illustration of the different currents in model A with $L=200$, $\beta=0.2$, $\Omega=0.1$, $\alpha=0.1$ and $\epsilon=1.55$. This choice of parameters corresponds to a domain wall position of $z\approx 0.88$ so that the TASEP occupancy is concentrated on a small region around the tip. The left panel shows a comparison of the current $J^{\mathrm{T}} (x)$ on the TASEP (filled red (dark grey) circles) and the currents $J^{\mathrm{S}}_{\mathrm{left}} (x)$ on the SSEP towards the base (left)  (filled orange (light grey) squares) and $J^{\mathrm{S}}_{\mathrm{right}} (x)$ towards the tip (right) (filled green (grey) diamonds). The TASEP current corresponds to the net SSEP current towards the base, that is the difference between the SSEP current towards the base and towards the tip. Both SSEP currents, however, are much larger than the TASEP current suggesting that the motors are mainly transported by the cytoplasm (SSEP) rather than by the filament (TASEP). The right panel shows both the attachment current $J_{\mathrm{on}} (x)$ (filled purple (dark grey) downward facing triangles) and the detachment current $J_{\mathrm{off}} (x)$ (light blue (light grey) upward facing triangles). Both currents are enhanced considerably towards the tip region. As a result, motor exchange happens primarily around the tip region. Furthermore, the maximum of the attachment current is located a little further away from the tip than the maximum of the detachment current so that typically motors attach to the filament, walk a short distance on the filament, detach near the tip and can then diffuse in the cytoplasm back to the cell body or reattach again. Note that the detachment current is particularly high at the last site due to the higher detachment rate at the tip. All currents are measured directly from the simulation that is counting the number of jumps per time.}
    \label{fig:allcurrentssupplement}
\end{figure}

Another point we want to emphasize is that in the system the nearest-neighbour correlations of the TASEP significantly modify the TASEP current in the sense that the mean-field prediction for it overestimates the current by orders of magnitude. The reason lies in the covariances that reach very high values of around $0.2$ in the region of the domain wall (see for example Fig. \ref{fig:DWcor}).  As we have shown above in Eq. (\ref{fisupplement}) the nearest-neighbour correlator $f_i = \left\langle n^{\mathrm{T}}_i n^{\mathrm{T}}_{i+1} \right\rangle$ is given by $f_i = \rho^{\mathrm{T}}_i + \omega \sum_{j=0}^i \left(\rho^{\mathrm{T}}_j - \rho^{\mathrm{S}}_j \right)$. Therefore, the covariances are given by
\begin{align*}
\mathrm{Cov}_i = \rho^{\mathrm{T}}_i (1-\rho^{\mathrm{T}}_{i+1}) +\omega \sum_{j=0}^i (\rho^{\mathrm{T}}_j - \rho^{\mathrm{S}}_j)=J^{\mathrm{T}}_{\mathrm{MF},i} +\omega \sum_{j=0}^i (\rho^{\mathrm{T}}_j - \rho^{\mathrm{S}}_j).
\end{align*}
Due to the first term, the mean-field current, which is large ($\approx 0.2$) in the region of the domain wall, also the covariances are non-zero and large there. However, since the mean-field current gives the main contribution to the covariances, this implies that the actual current 
\begin{align}
J^{\mathrm{T}}_i = \rho^{\mathrm{T}}_i - f_i = \omega \sum_{j=0}^i \left(\rho^{\mathrm{S}}_j - \rho^{\mathrm{T}}_j \right)=J^{\mathrm{T}}_{\mathrm{MF},i}-\mathrm{Cov}_i \label{actualcurrent}
\end{align}
is much smaller than predicted from a mean-field theory. To illustrate this further, in Fig. \ref{fig:comparecurrentsupplement} we show a comparison of the mean-field current on the left as obtained from simulations using the average densities, and of the actual current on the right as obtained directly from the simulation (filled red (dark grey) circles), from the average densities and nearest-neighbour correlations in the simulation (filled orange (grey) circles) and from Eq. (\ref{actualcurrent}) (dotted green (light grey) curve). Certainly, this discrepancy scales with $\Omega$ (and the other parameters) and the case shown here is extreme as $\Omega=0.001$ was chosen. However, the fact that both currents differ by orders of magnitude is robust and also occurs for $\Omega$ being of the order of the other jump rates [Fig.\ \ref{fig:allcurrentssupplement}]. This shows that a mean-field description fails to capture essential properties of our model system.

Furthermore, from the fact that the covariances are non-zero everywhere $\mathrm{Cov}_i \geq 0$ and that we can rewrite 
\begin{align*}
\mathrm{Cov}_i &= \left\langle n^{\mathrm{T}}_i n^{\mathrm{T}}_{i+1} \right\rangle - \rho^{\mathrm{T}}_i \rho^{\mathrm{T}}_{i+1} = \mathrm{Prob} \{ n^{\mathrm{T}}_i, n^{\mathrm{T}}_{i+1} \} - \mathrm{Prob} \{ n^{\mathrm{T}}_i\} \mathrm{Prob} \{n^{\mathrm{T}}_{i+1} \} = \\
&= \mathrm{Prob} \{ n^{\mathrm{T}}_i| n^{\mathrm{T}}_{i+1}  \} \mathrm{Prob} \{ n^{\mathrm{T}}_{i+1} \} - \mathrm{Prob} \{ n^{\mathrm{T}}_i\} \mathrm{Prob} \{ n^{\mathrm{T}}_{i+1} \}
\end{align*} 
we can conclude that $\mathrm{Prob} \{ n^{\mathrm{T}}_i| n^{\mathrm{T}}_{i+1}  \} \geq \mathrm{Prob} \{ n^{\mathrm{T}}_i\}$.  And from using a similar argument and only interchanging the roles of $i$ and $i+1$ we find $\mathrm{Prob} \{ n^{\mathrm{T}}_{i+1}| n^{\mathrm{T}}_{i}  \} \geq \mathrm{Prob} \{ n^{\mathrm{T}}_{i+1}\}$. This tells us that, in particular from the domain wall region onwards where covariances are high, the motors preferentially cluster and hinder each other effectively so that the mean time a particle spends at a certain site is increased considerably with respect to freely moving motors. Eventually, this leads to the substantial difference between the actual current and the mean-field current.

Potentially, those results are also important from a biological point of view. In case of an enhanced detachment rate at the filament's tip (model A), tip localisation is facilitated [Fig.\ \ref{fig:phasediagram}] and large traffic jams can be avoided. In this case there is only high filament density in a small region around the tip so that motors spend more time in the tip region than in the other part of the filament, a feature that might be favoured biologically as then the motors or their cargo might have more time to perform the necessary tasks at the tip. Furthermore, transport to the tip might be strongly promoted by diffusion in the cytoplasm (SSEP) whose currents in both the direction of the tip and of the base significantly exceed the filament (TASEP) current [Fig.\ \ref{fig:allcurrentssupplement}]. Certainly, in case of our original model where exclusion in the cytoplasm occurs at occupancy 1, the motors cannot really bypass each other, so that the motors do not circulate very often. This, however, should be greatly enhanced in the generalised model with carrying capacity $N_{\mathrm{max}}>1$ (see next chapter) or in a real biological system where the motors can overtake each other in the cytoplasm. The fact that the filament current is strongly suppressed by excluded volume effects might also be beneficial from a biological point of view as every motor step on the filament consumes ATP contrary to diffusion in the cytoplasm. Thus, transport of motors by the cytoplasm rather than by the filament might be advantageous energetically. Moreover, both the attachment and detachment current $J_{\mathrm{on}}$ and $J_{\mathrm{off}}$ are mainly restricted to the tip area [Fig.\ \ref{fig:allcurrentssupplement}] so that motor exchange between the filament and the cytoplasm occurs primarily near the tip where the cargo is used. Taken together, tip localisation and the suppression of the filament current might be beneficial from a biological point of view as then energy consumption is low and motors could be efficiently transported to the tip by the cytoplasm. Near the tip they attach to the filament, have an enhanced residence time on the filament due to steric hindrance between the motors and then detach at the tip back into the cytoplasm.

\section{Cylindrical geometry with several lanes for diffusion}

\begin{figure}[t]
    \centering
    \includegraphics[width=8.cm,valign=c]{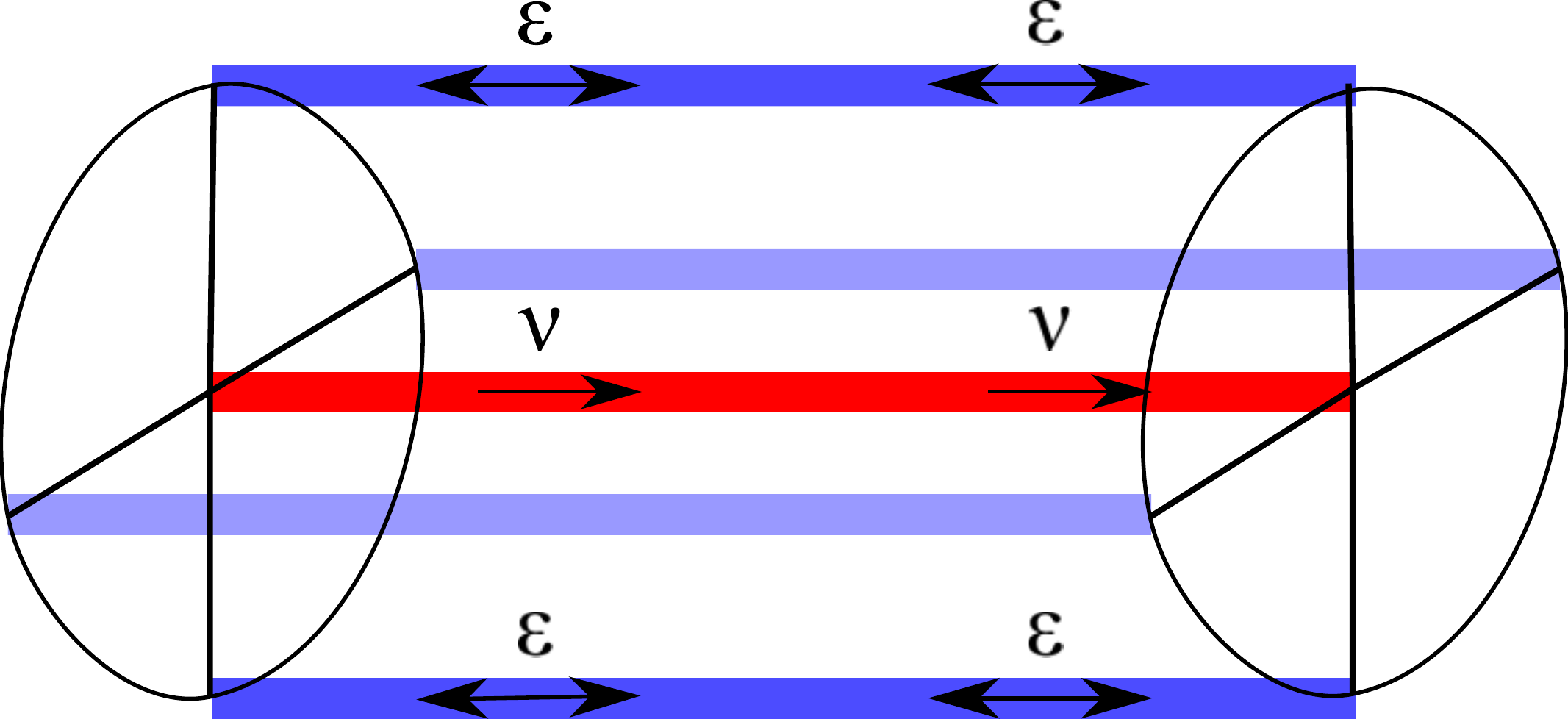} 
    \hfill
    \includegraphics[width=7.cm,valign=c]{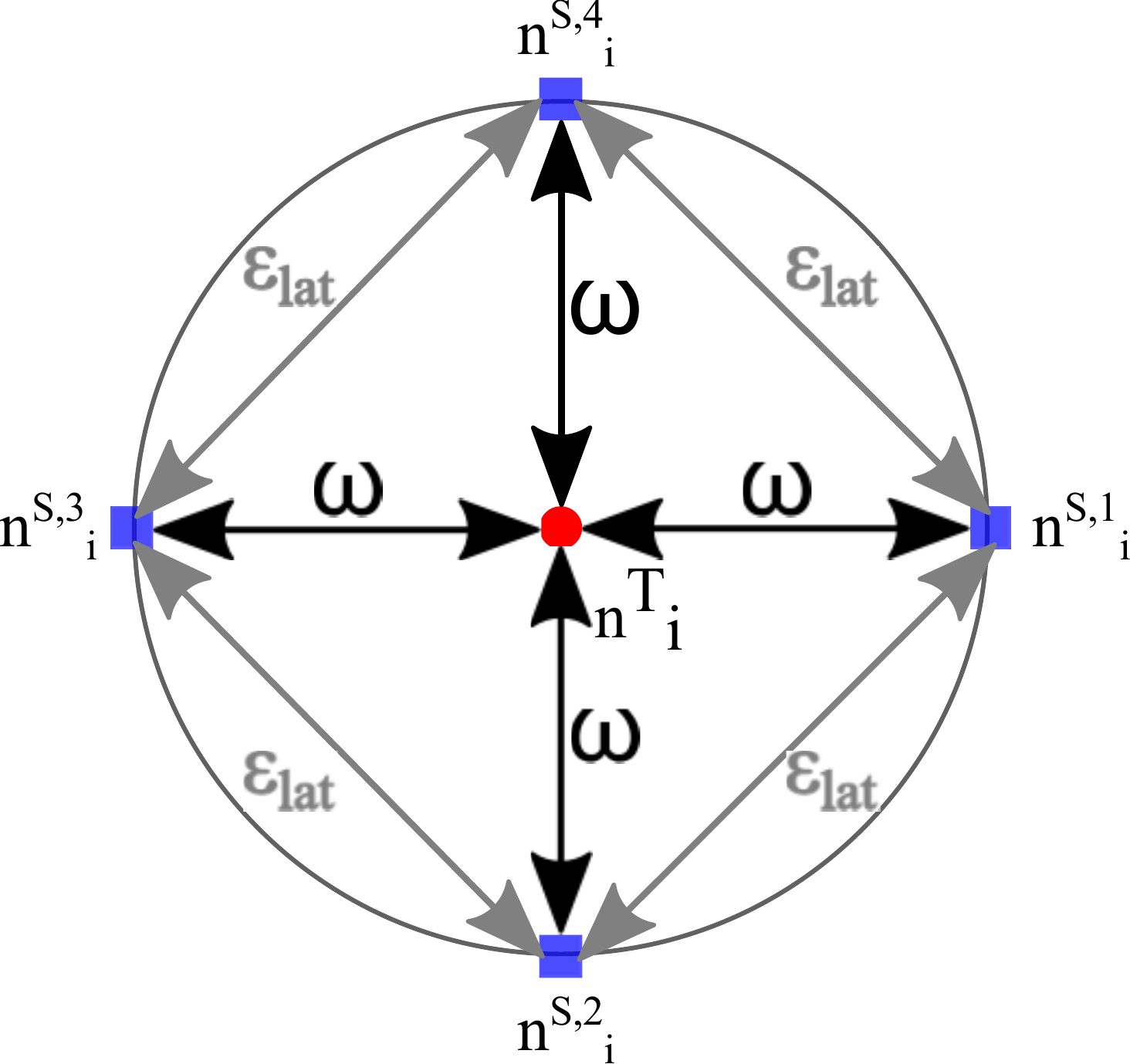} 
    \caption{Illustration of the generalised model with several (here: 4) lanes for diffusion: the three-dimensional view is shown on the left-hand side, the profile is illustrated on the right-hand side. As before the dynamics on the filament (red) are given by a TASEP with jump rate $\nu=1$ but now the dynamics in the cytoplasm is modelled by several lanes for diffusion (blue), each with diffusive rate $\epsilon$ and respecting the exclusion. The lanes for diffusion are arranged in a cylinder-like fashion around the TASEP and each can interact with the TASEP by attachment/detachment processes at rate $\omega$ respecting the exclusion. Apart from the diffusion along the cylinder axis there is also lateral diffusion between neighbouring lanes of diffusion. This diffusion happens at rate $\epsilon_{\mathrm{lat}}$ and again respects the exclusion. At the base (not shown) there is influx at rate $\alpha$ into every lane for diffusion and outflux at the diffusive rate $\epsilon$.}
    \label{fig:several_lanes}
\end{figure}

Finally, we want to deal with a generalisation of our model where instead of one lane for diffusion we have several lanes for diffusion arranged on a cylinder around the filament. For an illustration of the case with $N_{\mathrm{diff}}=4$ lanes for diffusion please refer to Fig. \ref{fig:several_lanes}. The three-dimensional view is shown on the left-hand side, the profile is illustrated on the right-hand side. As before the dynamics on the filament (red) are given by a TASEP lane with jump rate $\nu=1$ but now the dynamics in the cytoplasm is modelled by several lanes for diffusion (blue), each with diffusive rate $\epsilon$ and respecting the exclusion. The lanes for diffusion are arranged in a cylinder-like fashion around the TASEP lane, and each can interact with the TASEP lane by attachment/detachment processes at rate $\omega$ respecting the exclusion. Apart from the diffusion along the cylinder axis there is also lateral diffusion between neighbouring lanes of diffusion. This diffusion happens at rate $\epsilon_{\textbf{lat}}$ and again respects the exclusion. At the base (not shown) there is influx at rate $\alpha$ into every lane for diffusion and outflux at the diffusive rate $\epsilon$. 

In the following, we will show that we can reduce the steady-state behaviour of this more elaborate model to the steady-state behaviour of our model by scaling the parameters for diffusion and attachment/detachment by the number of lanes for diffusion $N_{\mathrm{diff}}$: $\omega \rightarrow \omega N_{\mathrm{diff}}$, $\epsilon \rightarrow \epsilon N_{\mathrm{diff}}$. Here, we only consider the case where we have the same attachment and detachment kinetics at the tip than in the bulk (model B), but an analogous generalisation and argument can be done for model A or different rates for attachment and detachment. Let us denote by $n^{\mathrm{T}}_i$ the occupancy at site $i$ of the TASEP (as before) and by $n^{\mathrm{S},m}_i$ the occupancy at site $i$ of the $m =1, \ldots, N_{\mathrm{diff}}$-th lane for diffusion. We start by $m=1$ at an arbitrary lane for diffusion and then consecutively number the lanes for diffusion in clockwise order. Since the lanes are arranged on a cylinder we have periodic boundary conditions and identify $n^{\mathrm{S},N_{\mathrm{diff}}+1}_i \equiv n^{\mathrm{S},1}_i$. With this convention we have the following bulk master equations, written straight away in terms of the averages over occupancies $\rho^{\mu,m}_i=\left\langle n^{\mu,m}_i \right\rangle$ and $f_i =\left\langle n^{\mathrm{T}}_i n^{\mathrm{T}}_{i+1} \right\rangle$:

\begin{align}
\partial_t \rho^{\mathrm{T}}_i &=  \rho^{\mathrm{T}}_{i-1}-f_{i-1}-\rho^{\mathrm{T}}_i + f_i + \sum_{m=1}^{N_{\mathrm{diff}}} \omega \left( \langle n^{\mathrm{S}, m}_i (1-n^{\mathrm{T}}_i) \rangle - \langle n^{\mathrm{T}}_i (1-n^{\mathrm{S},m}_i) \rangle \right) \label{eq:several_lanes_bulk} \\
\partial_t \rho^{\mathrm{S},m}_i &= \epsilon \left( \rho^{\mathrm{S},m}_{i+1}+\rho^{\mathrm{S},m}_{i-1} - 2 \rho^{\mathrm{S},m}_i \right) + \omega \left( \langle n^{\mathrm{T}}_i (1-n^{\mathrm{S},m}_i) \rangle - \langle n^{\mathrm{S},m}_i (1-n^{\mathrm{T}}_i) \rangle \right) + \epsilon_{\mathrm{lat}} \left(\rho^{\mathrm{S}, m+1}_i + \rho^{\mathrm{S},m-1}_i - 2 \rho^{\mathrm{S},m}_i \right). \nonumber
\end{align}
At the base we find

\begin{align}
\partial_t \rho^{\mathrm{T}}_0 &=  -\rho^{\mathrm{T}}_0 + f_0 + \sum_{m=1}^{N_{\mathrm{diff}}} \omega \left( \langle n^{\mathrm{S}, m}_0 (1-n^{\mathrm{T}}_0) \rangle - \langle n^{\mathrm{T}}_0 (1-n^{\mathrm{S},m}_0) \rangle \right) \label{eq:several_lanes_base} \\
\partial_t \rho^{\mathrm{S},m}_0 &= \alpha -(\alpha+\epsilon) \rho^{\mathrm{S},m}_0 + \epsilon \left( \rho^{\mathrm{S},m}_{1}- \rho^{\mathrm{S},m}_0 \right) + \omega \left( \langle n^{\mathrm{T}}_0 (1-n^{\mathrm{S},m}_0) \rangle - \langle n^{\mathrm{S},m}_0 (1-n^{\mathrm{T}}_0) \rangle \right) + \epsilon_{\mathrm{lat}} \left(\rho^{\mathrm{S}, m+1}_0 + \rho^{\mathrm{S},m-1}_0 - 2 \rho^{\mathrm{S},m}_0 \right) \nonumber
\end{align}
and at the tip

\begin{align}
\partial_t \rho^{\mathrm{T}}_L &=  \rho^{\mathrm{T}}_{L-1}-f_{L-1}+ \sum_{m=1}^{N_{\mathrm{diff}}} \omega \left( \langle n^{\mathrm{S}, m}_L (1-n^{\mathrm{T}}_L) \rangle - \langle n^{\mathrm{T}}_L (1-n^{\mathrm{S},m}_L) \rangle \right) \label{eq:several_lanes_tip}\\
\partial_t \rho^{\mathrm{S},m}_L &= \epsilon \left( \rho^{\mathrm{S},m}_{L-1} -  \rho^{\mathrm{S},m}_L \right) + \omega \left( \langle n^{\mathrm{T}}_L (1-n^{\mathrm{S},m}_L) \rangle - \langle n^{\mathrm{S},m}_L (1-n^{\mathrm{T}}_L) \rangle \right) + \epsilon_{\mathrm{lat}} \left(\rho^{\mathrm{S}, m+1}_L + \rho^{\mathrm{S},m-1}_L - 2 \rho^{\mathrm{S},m}_L \right). \nonumber
\end{align}
Due to the exclusion property on both the TASEP as well as on the lanes for diffusion the state space of the system is finite. Furthermore, it is an irreducible continuous-time Markov process so there exists a unique steady-state. This is important since then the cylindrical symmetry of the system must be reflected in this steady-state. This is why we can assume that in the steady-state $\left\langle n^{\mathrm{S},m}_i \right\rangle=\left\langle n^{\mathrm{S},n}_i \right\rangle$ and $\langle n^{\mathrm{S}, m}_i n^{\mathrm{T}}_i \rangle=\langle n^{\mathrm{S}, n}_i n^{\mathrm{T}}_i \rangle$ holds for all $m,n$. As a result, the terms proportional to $\epsilon_{\mathrm{lat}}$ drop out in Eqs. (\ref{eq:several_lanes_bulk})-(\ref{eq:several_lanes_tip}) and we can define the total occupancy at site $i$ in the cytoplasm

\begin{align*}
n^{\mathrm{S}}_i \equiv \sum_{m=1}^{N_{\mathrm{diff}}} n^{\mathrm{S},m}_i
\end{align*}
and the corresponding average $\rho^{\mathrm{S}}_i$ as the sum of the (average) occupancies at sites $i$ of all the lanes for diffusion taken together. Using this quantity we can rewrite Eqs. (\ref{eq:several_lanes_bulk})-(\ref{eq:several_lanes_tip}) in steady-state as follows:

\begin{align}
0 &=  \rho^{\mathrm{T}}_{i-1}-f_{i-1}-\rho^{\mathrm{T}}_i + f_i +  \omega \left( \langle n^{\mathrm{S}}_i (1-n^{\mathrm{T}}_i) \rangle - \langle n^{\mathrm{T}}_i (N_{\mathrm{diff}}-n^{\mathrm{S}}_i) \rangle \right) \label{eq:several_lanes_bulk_total_occ} \\
0 &= \epsilon \left( \rho^{\mathrm{S}}_{i+1}+\rho^{\mathrm{S}}_{i-1} - 2 \rho^{\mathrm{S}}_i \right) + \omega \left( \langle n^{\mathrm{T}}_i (N_{\mathrm{diff}}-n^{\mathrm{S}}_i) \rangle - \langle n^{\mathrm{S}}_i (1-n^{\mathrm{T}}_i) \rangle \right) \nonumber
\end{align}
for the bulk master equation,

\begin{align}
0 &=  -\rho^{\mathrm{T}}_0 + f_0 +  \omega \left( \langle n^{\mathrm{S}}_0 (1-n^{\mathrm{T}}_0) \rangle - \langle n^{\mathrm{T}}_0 (N_{\mathrm{diff}}-n^{\mathrm{S}}_0) \rangle \right) \label{eq:several_lanes_base_total_occ} \\
0 &= \alpha  N_{\mathrm{diff}} -(\alpha+\epsilon) \rho^{\mathrm{S}}_0 + \epsilon \left( \rho^{\mathrm{S}}_{1}- \rho^{\mathrm{S}}_0 \right) + \omega \left( \langle n^{\mathrm{T}}_0 (N_{\mathrm{diff}}-n^{\mathrm{S}}_0) \rangle - \langle n^{\mathrm{S}}_0 (1-n^{\mathrm{T}}_0) \rangle \right) \nonumber
\end{align}
for the base and

\begin{align}
0 &=  \rho^{\mathrm{T}}_{L-1}-f_{L-1}+  \omega \left( \langle n^{\mathrm{S}}_L (1-n^{\mathrm{T}}_L) \rangle - \langle n^{\mathrm{T}}_L (N_{\mathrm{diff}}-n^{\mathrm{S}}_L) \rangle \right) \label{eq:several_lanes_tip_total_occ}\\
0 &= \epsilon \left( \rho^{\mathrm{S}}_{L-1} -  \rho^{\mathrm{S}}_L \right) + \omega \left( \langle n^{\mathrm{T}}_L (N_{\mathrm{diff}}-n^{\mathrm{S}}_L) \rangle - \langle n^{\mathrm{S}}_L (1-n^{\mathrm{T}}_L) \rangle \right) \nonumber
\end{align}
for the tip. From these equations it becomes apparent that the cylindrical system with $N_{\mathrm{diff}}$ lanes for diffusion is equivalent to the above mentioned generalisation of our model where exclusion on the filament does not happen at occupancy of 1 but at a maximal occupancy or carrying capacity of $N_{\mathrm{max}} = N_{\mathrm{diff}}$ that is reflected in the term $N_{\mathrm{diff}} - n^{\mathrm{S}}_i$. To show that the steady-state behaviour of those two generalisations of our model are cast by the steady-state behaviour of our model, we next introduce the quantity
\begin{align*}
\tilde{n}^{\mathrm{S}}_i = \frac{1}{N_\mathrm{diff}} n^{\mathrm{S}}_i = \frac{1}{N_{\mathrm{diff}}} \sum_{m=1}^{N_{\mathrm{diff}}} n^{\mathrm{S},m}_i
\end{align*}
and the corresponding average occupancy $\tilde{\rho}^{\mathrm{S}}_i$ at site $i$ of one lane for diffusion. In terms of $\tilde{n}^{\mathrm{S}}_i$ we find

\begin{align}
0 &=  \rho^{\mathrm{T}}_{i-1}-f_{i-1}-\rho^{\mathrm{T}}_i + f_i +  \omega N_{\mathrm{diff}} \left( \tilde{\rho}^{\mathrm{S}}_i - \rho^{\mathrm{T}}_i \right) \label{eq:several_lanes_bulk_av_occ} \\
0 &= \epsilon N_{\mathrm{diff}} \left( \tilde{\rho}^{\mathrm{S}}_{i+1}+\tilde{\rho}^{\mathrm{S}}_{i-1} - 2 \tilde{\rho}^{\mathrm{S}}_i \right) + \omega N_{\mathrm{diff}} \left( \rho^{\mathrm{T}}_i - \tilde{\rho}^{\mathrm{S}}_i\right) \nonumber
\end{align}
for the bulk and

\begin{align}
0 &=  -\rho^{\mathrm{T}}_0 + f_0 +  \omega N_{\mathrm{diff}} \left( \tilde{\rho}^{\mathrm{S}}_0 - \rho^{\mathrm{T}}_0\right) \label{eq:several_lanes_base_av_occ} \\
0 &= \alpha N_{\mathrm{diff}}  -(\alpha+\epsilon) N_{\mathrm{diff}} \tilde{\rho}^{\mathrm{S}}_0 + \epsilon N_{\mathrm{diff}} \left( \tilde{\rho}^{\mathrm{S}}_{1}- \tilde{\rho}^{\mathrm{S}}_0 \right) + \omega N_{\mathrm{diff}} \left( \rho^{\mathrm{T}}_0 - \tilde{\rho}^{\mathrm{S}}_0 \right) \nonumber\\
0 &=  \rho^{\mathrm{T}}_{L-1}-f_{L-1}+  \omega N_{\mathrm{diff}} \left( \tilde{\rho}^{\mathrm{S}}_L - \rho^{\mathrm{T}}_L \right) \label{eq:several_lanes_tip_av_occ}\\
0 &= \epsilon N_{\mathrm{diff}} \left( \tilde{\rho}^{\mathrm{S}}_{L-1} -  \tilde{\rho}^{\mathrm{S}}_L \right) + \omega N_{\mathrm{diff}} \left( \rho^{\mathrm{T}}_L - \tilde{\rho}^{\mathrm{S}}_L \right) \nonumber
\end{align}
for the base and the tip where we used that $ \langle \tilde{n}^{\mathrm{S}}_i (1-n^{\mathrm{T}}_i) \rangle - \langle n^{\mathrm{T}}_i (1-\tilde{n}^{\mathrm{S}}_i) \rangle = \tilde{\rho}^{\mathrm{S}}_i - \rho^{\mathrm{T}}_i$. \newline
In summary, if we replace $\omega\rightarrow \omega N_{\mathrm{diff}}$, $\epsilon \rightarrow \epsilon N_{\mathrm{diff}}$ and $\alpha \rightarrow \alpha N_{\mathrm{diff}}$ in our model we can deduce from the steady-state behaviour of our model the steady-state behaviour of these generalised models. Certainly, to stay within the scope of our considerations, this implies that we need $\omega N_{\mathrm{diff}}$ to be much smaller than 1 (order $a$) and $\epsilon N_{\mathrm{diff}}$ to be of the order of 1 implying that the "total attachment/detachment rate`` $\Omega N_{\mathrm{diff}}$ and the "total diffusion rate`` $\epsilon N_{\mathrm{diff}}$ should be of the order of the hopping constant $\nu =1$.


\begin{thebibliography}{53}%
\makeatletter
\providecommand \@ifxundefined [1]{%
 \@ifx{#1\undefined}
}%
\providecommand \@ifnum [1]{%
 \ifnum #1\expandafter \@firstoftwo
 \else \expandafter \@secondoftwo
 \fi
}%
\providecommand \@ifx [1]{%
 \ifx #1\expandafter \@firstoftwo
 \else \expandafter \@secondoftwo
 \fi
}%
\providecommand \natexlab [1]{#1}%
\providecommand \enquote  [1]{``#1''}%
\providecommand \bibnamefont  [1]{#1}%
\providecommand \bibfnamefont [1]{#1}%
\providecommand \citenamefont [1]{#1}%
\providecommand \href@noop [0]{\@secondoftwo}%
\providecommand \href [0]{\begingroup \@sanitize@url \@href}%
\providecommand \@href[1]{\@@startlink{#1}\@@href}%
\providecommand \@@href[1]{\endgroup#1\@@endlink}%
\providecommand \@sanitize@url [0]{\catcode `\\12\catcode `\$12\catcode
  `\&12\catcode `\#12\catcode `\^12\catcode `\_12\catcode `\%12\relax}%
\providecommand \@@startlink[1]{}%
\providecommand \@@endlink[0]{}%
\providecommand \url  [0]{\begingroup\@sanitize@url \@url }%
\providecommand \@url [1]{\endgroup\@href {#1}{\urlprefix }}%
\providecommand \urlprefix  [0]{URL }%
\providecommand \Eprint [0]{\href }%
\providecommand \doibase [0]{http://dx.doi.org/}%
\providecommand \selectlanguage [0]{\@gobble}%
\providecommand \bibinfo  [0]{\@secondoftwo}%
\providecommand \bibfield  [0]{\@secondoftwo}%
\providecommand \translation [1]{[#1]}%
\providecommand \BibitemOpen [0]{}%
\providecommand \bibitemStop [0]{}%
\providecommand \bibitemNoStop [0]{.\EOS\space}%
\providecommand \EOS [0]{\spacefactor3000\relax}%
\providecommand \BibitemShut  [1]{\csname bibitem#1\endcsname}%
\let\auto@bib@innerbib\@empty
\bibitem [{\citenamefont {Revenu}\ \emph {et~al.}(2004)\citenamefont {Revenu},
  \citenamefont {Athman}, \citenamefont {Robine},\ and\ \citenamefont
  {Louvard}}]{Revenu:2004aa}%
  \BibitemOpen
  \bibfield  {author} {\bibinfo {author} {\bibfnamefont {C.}~\bibnamefont
  {Revenu}}, \bibinfo {author} {\bibfnamefont {R.}~\bibnamefont {Athman}},
  \bibinfo {author} {\bibfnamefont {S.}~\bibnamefont {Robine}}, \ and\ \bibinfo
  {author} {\bibfnamefont {D.}~\bibnamefont {Louvard}},\ }\href
  {http://dx.doi.org/10.1038/nrm1437} {\bibfield  {journal} {\bibinfo
  {journal} {Nat. Rev. Mol. Cell Biol.}\ }\textbf {\bibinfo {volume} {5}},\
  \bibinfo {pages} {635} (\bibinfo {year} {2004})}\BibitemShut {NoStop}%
\bibitem [{\citenamefont {Mattila}\ and\ \citenamefont
  {Lappalainen}(2008)}]{Mattila:2008aa}%
  \BibitemOpen
  \bibfield  {author} {\bibinfo {author} {\bibfnamefont {P.~K.}\ \bibnamefont
  {Mattila}}\ and\ \bibinfo {author} {\bibfnamefont {P.}~\bibnamefont
  {Lappalainen}},\ }\href {http://dx.doi.org/10.1038/nrm2406} {\bibfield
  {journal} {\bibinfo  {journal} {Nat. Rev. Mol. Cell Biol.}\ }\textbf
  {\bibinfo {volume} {9}},\ \bibinfo {pages} {446} (\bibinfo {year}
  {2008})}\BibitemShut {NoStop}%
\bibitem [{\citenamefont {Les~Erickson}\ \emph {et~al.}(2003)\citenamefont
  {Les~Erickson}, \citenamefont {Corsa}, \citenamefont {Dos{\'e}},\ and\
  \citenamefont {Burnside}}]{LesErickson01102003}%
  \BibitemOpen
  \bibfield  {author} {\bibinfo {author} {\bibfnamefont {F.}~\bibnamefont
  {Les~Erickson}}, \bibinfo {author} {\bibfnamefont {A.~C.}\ \bibnamefont
  {Corsa}}, \bibinfo {author} {\bibfnamefont {A.~C.}\ \bibnamefont {Dos{\'e}}},
  \ and\ \bibinfo {author} {\bibfnamefont {B.}~\bibnamefont {Burnside}},\
  }\href {\doibase 10.1091/mbc.E02-10-0656} {\bibfield  {journal} {\bibinfo
  {journal} {Mol. Biol. Cell}\ }\textbf {\bibinfo {volume} {14}},\ \bibinfo
  {pages} {4173} (\bibinfo {year} {2003})}\BibitemShut {NoStop}%
\bibitem [{\citenamefont {Nambiar}\ \emph {et~al.}(2010)\citenamefont
  {Nambiar}, \citenamefont {McConnell},\ and\ \citenamefont
  {Tyska}}]{Nambiar2010}%
  \BibitemOpen
  \bibfield  {author} {\bibinfo {author} {\bibfnamefont {R.}~\bibnamefont
  {Nambiar}}, \bibinfo {author} {\bibfnamefont {R.~E.}\ \bibnamefont
  {McConnell}}, \ and\ \bibinfo {author} {\bibfnamefont {M.~J.}\ \bibnamefont
  {Tyska}},\ }\href {\doibase 10.1007/s00018-009-0254-5} {\bibfield  {journal}
  {\bibinfo  {journal} {Cell. Mol. Life Sci.}\ }\textbf {\bibinfo {volume}
  {67}},\ \bibinfo {pages} {1239} (\bibinfo {year} {2010})}\BibitemShut
  {NoStop}%
\bibitem [{\citenamefont {Salles}\ \emph {et~al.}(2009)\citenamefont {Salles},  \citenamefont {Merritt}, 
  \citenamefont{Manor}, \citenamefont {Dougherty},  \citenamefont {Sousa}, \citenamefont {Moore}, \citenamefont {Yengo}, 
  \citenamefont {Dose},\ and\ \citenamefont {Kachar}}]{Salles:2009aa}%
  \BibitemOpen
  \bibfield  {author} {\bibinfo {author} {\bibfnamefont {F.~T.}\ \bibnamefont
  {Salles}}, \bibinfo {author} {\bibfnamefont {R.~C.}\ \bibnamefont {Merritt}},
  \bibinfo {author} {\bibfnamefont {U.}~\bibnamefont {Manor}}, \bibinfo
  {author} {\bibfnamefont {G.~W.}\ \bibnamefont {Dougherty}}, \bibinfo {author}
  {\bibfnamefont {A.~D.}\ \bibnamefont {Sousa}}, \bibinfo {author}
  {\bibfnamefont {J.~E.}\ \bibnamefont {Moore}}, \bibinfo {author}
  {\bibfnamefont {C.~M.}\ \bibnamefont {Yengo}}, \bibinfo {author}
  {\bibfnamefont {A.~C.}\ \bibnamefont {Dose}}, \ and\ \bibinfo {author}
  {\bibfnamefont {B.}~\bibnamefont {Kachar}},\ }\href
  {http://dx.doi.org/10.1038/ncb1851} {\bibfield  {journal} {\bibinfo
  {journal} {Nat. Cell Biol.}\ }\textbf {\bibinfo {volume} {11}},\ \bibinfo
  {pages} {443} (\bibinfo {year} {2009})}\BibitemShut {NoStop}%
\bibitem [{\citenamefont {Hartman}\ and\ \citenamefont
  {Spudich}(2012)}]{Hartman1627}%
  \BibitemOpen
  \bibfield  {author} {\bibinfo {author} {\bibfnamefont {M.~A.}\ \bibnamefont
  {Hartman}}\ and\ \bibinfo {author} {\bibfnamefont {J.~A.}\ \bibnamefont
  {Spudich}},\ }\href {\doibase 10.1242/jcs.094300} {\bibfield  {journal}
  {\bibinfo  {journal} {J. Cell Sci.}\ }\textbf {\bibinfo {volume} {125}},\
  \bibinfo {pages} {1627} (\bibinfo {year} {2012})} \BibitemShut
  {NoStop}%
\bibitem [{\citenamefont {Kerber}\ and\ \citenamefont
  {Cheney}(2011)}]{Kerber3733}%
  \BibitemOpen
  \bibfield  {author} {\bibinfo {author} {\bibfnamefont {M.~L.}\ \bibnamefont
  {Kerber}}\ and\ \bibinfo {author} {\bibfnamefont {R.~E.}\ \bibnamefont
  {Cheney}},\ }\href {\doibase 10.1242/jcs.023549} {\bibfield  {journal}
  {\bibinfo  {journal} {J. Cell Sci.}\ }\textbf {\bibinfo {volume} {124}},\
  \bibinfo {pages} {3733} (\bibinfo {year} {2011})} \BibitemShut
  {NoStop}%
\bibitem [{\citenamefont {Kambara}\ \emph {et~al.}(2006)\citenamefont
  {Kambara}, \citenamefont {Komaba},\ and\ \citenamefont
  {Ikebe}}]{Kambara08122006}%
  \BibitemOpen
  \bibfield  {author} {\bibinfo {author} {\bibfnamefont {T.}~\bibnamefont
  {Kambara}}, \bibinfo {author} {\bibfnamefont {S.}~\bibnamefont {Komaba}}, \
  and\ \bibinfo {author} {\bibfnamefont {M.}~\bibnamefont {Ikebe}},\ }\href
  {\doibase 10.1074/jbc.M603823200} {\bibfield  {journal} {\bibinfo  {journal}
  {J. Biol. Chem.}\ }\textbf {\bibinfo {volume} {281}},\ \bibinfo {pages}
  {37291} (\bibinfo {year} {2006})}\BibitemShut
  {NoStop}%
\bibitem [{\citenamefont {Bird}\ \emph {et~al.}(2014)\citenamefont {Bird},
  \citenamefont {Takagi}, \citenamefont {Billington}, \citenamefont {Strub},
  \citenamefont {Sellers},\ and\ \citenamefont {Friedman}}]{Bird26082014}%
  \BibitemOpen
  \bibfield  {author} {\bibinfo {author} {\bibfnamefont {J.~E.}\ \bibnamefont
  {Bird}}, \bibinfo {author} {\bibfnamefont {Y.}~\bibnamefont {Takagi}},
  \bibinfo {author} {\bibfnamefont {N.}~\bibnamefont {Billington}}, \bibinfo
  {author} {\bibfnamefont {M.-P.}\ \bibnamefont {Strub}}, \bibinfo {author}
  {\bibfnamefont {J.~R.}\ \bibnamefont {Sellers}}, \ and\ \bibinfo {author}
  {\bibfnamefont {T.~B.}\ \bibnamefont {Friedman}},\ }\href {\doibase
  10.1073/pnas.1409459111} {\bibfield  {journal} {\bibinfo  {journal} {Proc.
  Nat. Acad. Sci. USA}\ }\textbf {\bibinfo {volume} {111}},\ \bibinfo {pages}
  {12390} (\bibinfo {year} {2014})}\BibitemShut {NoStop}%
\bibitem [{\citenamefont {Rzadzinska}\ \emph {et~al.}(2004)\citenamefont
  {Rzadzinska}, \citenamefont {Schneider}, \citenamefont {Davies},
  \citenamefont {Riordan},\ and\ \citenamefont {Kachar}}]{Rzadzinska15032004}%
  \BibitemOpen
  \bibfield  {author} {\bibinfo {author} {\bibfnamefont {A.~K.}\ \bibnamefont
  {Rzadzinska}}, \bibinfo {author} {\bibfnamefont {M.~E.}\ \bibnamefont
  {Schneider}}, \bibinfo {author} {\bibfnamefont {C.}~\bibnamefont {Davies}},
  \bibinfo {author} {\bibfnamefont {G.~P.}\ \bibnamefont {Riordan}}, \ and\
  \bibinfo {author} {\bibfnamefont {B.}~\bibnamefont {Kachar}},\ }\href
  {\doibase 10.1083/jcb.200310055} {\bibfield  {journal} {\bibinfo  {journal}
  {J. Cell Biol.}\ }\textbf {\bibinfo {volume} {164}},\ \bibinfo {pages} {887}
  (\bibinfo {year} {2004})}\BibitemShut
  {NoStop}%
\bibitem [{\citenamefont {Manor}\ \emph {et~al.}(2011)\citenamefont {Manor},
  \citenamefont {Disanza}, \citenamefont {Grati}, \citenamefont {Andrade},
  \citenamefont {Lin}, \citenamefont {Fiore}, \citenamefont {Scita},\ and\
  \citenamefont {Kachar}}]{Manor2011167}%
  \BibitemOpen
  \bibfield  {author} {\bibinfo {author} {\bibfnamefont {U.}~\bibnamefont
  {Manor}}, \bibinfo {author} {\bibfnamefont {A.}~\bibnamefont {Disanza}},
  \bibinfo {author} {\bibfnamefont {M.}~\bibnamefont {Grati}}, \bibinfo
  {author} {\bibfnamefont {L.}~\bibnamefont {Andrade}}, \bibinfo {author}
  {\bibfnamefont {H.}~\bibnamefont {Lin}}, \bibinfo {author} {\bibfnamefont
  {P.~P.~D.}\ \bibnamefont {Fiore}}, \bibinfo {author} {\bibfnamefont
  {G.}~\bibnamefont {Scita}}, \ and\ \bibinfo {author} {\bibfnamefont
  {B.}~\bibnamefont {Kachar}},\ }\href {\doibase
  http://dx.doi.org/10.1016/j.cub.2010.12.046} {\bibfield  {journal} {\bibinfo
  {journal} {Curr. Biol.}\ }\textbf {\bibinfo {volume} {21}},\ \bibinfo {pages}
  {167 } (\bibinfo {year} {2011})}\BibitemShut {NoStop}%
\bibitem [{\citenamefont {Belyantseva}\ \emph {et~al.}(2003)\citenamefont
  {Belyantseva}, \citenamefont {Boger},\ and\ \citenamefont
  {Friedman}}]{Belyantseva25112003}%
  \BibitemOpen
  \bibfield  {author} {\bibinfo {author} {\bibfnamefont {I.~A.}\ \bibnamefont
  {Belyantseva}}, \bibinfo {author} {\bibfnamefont {E.~T.}\ \bibnamefont
  {Boger}}, \ and\ \bibinfo {author} {\bibfnamefont {T.~B.}\ \bibnamefont
  {Friedman}},\ }\href {\doibase 10.1073/pnas.2334417100} {\bibfield  {journal}
  {\bibinfo  {journal} {Proc. Nat. Acad. Sci. USA}\ }\textbf {\bibinfo {volume}
  {100}},\ \bibinfo {pages} {13958} (\bibinfo {year} {2003})} \BibitemShut {NoStop}%
\bibitem [{\citenamefont {Zhuravlev}\ \emph {et~al.}(2012)\citenamefont
  {Zhuravlev}, \citenamefont {Lan}, \citenamefont {Minakova},\ and\
  \citenamefont {Papoian}}]{zhuravlev2}%
  \BibitemOpen
  \bibfield  {author} {\bibinfo {author} {\bibfnamefont {P.}~\bibnamefont
  {Zhuravlev}}, \bibinfo {author} {\bibfnamefont {Y.}~\bibnamefont {Lan}},
  \bibinfo {author} {\bibfnamefont {M.}~\bibnamefont {Minakova}}, \ and\
  \bibinfo {author} {\bibfnamefont {G.}~\bibnamefont {Papoian}},\ }\href
  {http://www.pnas.org/content/109/27/10849.abstract} {\bibfield  {journal}
  {\bibinfo  {journal} {Proc. Nat. Acad. Sci. USA}\ }\textbf {\bibinfo {volume}
  {109}},\ \bibinfo {pages} {10849} (\bibinfo {year} {2012})}\BibitemShut
  {NoStop}%
\bibitem [{\citenamefont {Naoz}\ \emph {et~al.}(2008)\citenamefont {Naoz},
  \citenamefont {Manor}, \citenamefont {Sakaguchi}, \citenamefont {Kachar},\
  and\ \citenamefont {Gov}}]{NaozGov}%
  \BibitemOpen
  \bibfield  {author} {\bibinfo {author} {\bibfnamefont {M.}~\bibnamefont
  {Naoz}}, \bibinfo {author} {\bibfnamefont {U.}~\bibnamefont {Manor}},
  \bibinfo {author} {\bibfnamefont {H.}~\bibnamefont {Sakaguchi}}, \bibinfo
  {author} {\bibfnamefont {B.}~\bibnamefont {Kachar}}, \ and\ \bibinfo {author}
  {\bibfnamefont {N.~S.}\ \bibnamefont {Gov}},\ }\href {\doibase
  10.1529/biophysj.108.143453} {\bibfield  {journal} {\bibinfo  {journal}
  {Biophys. J.}\ }\textbf {\bibinfo {volume} {95}},\ \bibinfo {pages} {5706}
  (\bibinfo {year} {2008})}\BibitemShut {NoStop}%
\bibitem [{\citenamefont {Wolff}\ \emph {et~al.}(2014)\citenamefont {Wolff},
  \citenamefont {Barrett-Freeman}, \citenamefont {Evans}, \citenamefont
  {Goryachev},\ and\ \citenamefont {Marenduzzo}}]{1478-3975-11-1-016005}%
  \BibitemOpen
  \bibfield  {author} {\bibinfo {author} {\bibfnamefont {K.}~\bibnamefont
  {Wolff}}, \bibinfo {author} {\bibfnamefont {C.}~\bibnamefont
  {Barrett-Freeman}}, \bibinfo {author} {\bibfnamefont {M.~R.}\ \bibnamefont
  {Evans}}, \bibinfo {author} {\bibfnamefont {A.~B.}\ \bibnamefont
  {Goryachev}}, \ and\ \bibinfo {author} {\bibfnamefont {D.}~\bibnamefont
  {Marenduzzo}},\ }\href {http://stacks.iop.org/1478-3975/11/i=1/a=016005}
  {\bibfield  {journal} {\bibinfo  {journal} {Phys. Biol.}\ }\textbf {\bibinfo
  {volume} {11}},\ \bibinfo {pages} {016005} (\bibinfo {year}
  {2014})}\BibitemShut {NoStop}%
\bibitem [{\citenamefont {Krug}(1991)}]{PhysRevLett.67.1882}%
  \BibitemOpen
  \bibfield  {author} {\bibinfo {author} {\bibfnamefont {J.}~\bibnamefont
  {Krug}},\ }\href {\doibase 10.1103/PhysRevLett.67.1882} {\bibfield  {journal}
  {\bibinfo  {journal} {Phys. Rev. Lett.}\ }\textbf {\bibinfo {volume} {67}},\
  \bibinfo {pages} {1882} (\bibinfo {year} {1991})}\BibitemShut {NoStop}%
\bibitem [{\citenamefont {Lipowsky}\ \emph {et~al.}(2001)\citenamefont
  {Lipowsky}, \citenamefont {Klumpp},\ and\ \citenamefont
  {Nieuwenhuizen}}]{lipowsky}%
  \BibitemOpen
  \bibfield  {author} {\bibinfo {author} {\bibfnamefont {R.}~\bibnamefont
  {Lipowsky}}, \bibinfo {author} {\bibfnamefont {S.}~\bibnamefont {Klumpp}}, \
  and\ \bibinfo {author} {\bibfnamefont {T.~M.}\ \bibnamefont
  {Nieuwenhuizen}},\ }\href@noop {} {\bibfield  {journal} {\bibinfo  {journal}
  {Phys. Rev. Lett.}\ }\textbf {\bibinfo {volume} {87}},\ \bibinfo {pages}
  {108101} (\bibinfo {year} {2001})}\BibitemShut {NoStop}%
\bibitem [{\citenamefont {Parmeggiani}\ \emph {et~al.}(2003)\citenamefont
  {Parmeggiani}, \citenamefont {Franosch},\ and\ \citenamefont
  {Frey}}]{PhysRevLett.90.086601}%
  \BibitemOpen
  \bibfield  {author} {\bibinfo {author} {\bibfnamefont {A.}~\bibnamefont
  {Parmeggiani}}, \bibinfo {author} {\bibfnamefont {T.}~\bibnamefont
  {Franosch}}, \ and\ \bibinfo {author} {\bibfnamefont {E.}~\bibnamefont
  {Frey}},\ }\href {\doibase 10.1103/PhysRevLett.90.086601} {\bibfield
  {journal} {\bibinfo  {journal} {Phys. Rev. Lett.}\ }\textbf {\bibinfo
  {volume} {90}},\ \bibinfo {pages} {086601} (\bibinfo {year}
  {2003})}\BibitemShut {NoStop}%
\bibitem [{\citenamefont {Parmeggiani}\ \emph {et~al.}(2004)\citenamefont
  {Parmeggiani}, \citenamefont {Franosch},\ and\ \citenamefont
  {Frey}}]{PhysRevE.70.046101}%
  \BibitemOpen
  \bibfield  {author} {\bibinfo {author} {\bibfnamefont {A.}~\bibnamefont
  {Parmeggiani}}, \bibinfo {author} {\bibfnamefont {T.}~\bibnamefont
  {Franosch}}, \ and\ \bibinfo {author} {\bibfnamefont {E.}~\bibnamefont
  {Frey}},\ }\href {\doibase 10.1103/PhysRevE.70.046101} {\bibfield  {journal}
  {\bibinfo  {journal} {Phys. Rev. E}\ }\textbf {\bibinfo {volume} {70}},\
  \bibinfo {pages} {046101} (\bibinfo {year} {2004})}\BibitemShut {NoStop}%
\bibitem [{\citenamefont {Evans}\ \emph {et~al.}(2011)\citenamefont {Evans},
  \citenamefont {Kafri}, \citenamefont {Sugden},\ and\ \citenamefont
  {Tailleur}}]{evans}%
  \BibitemOpen
  \bibfield  {author} {\bibinfo {author} {\bibfnamefont {M.}~\bibnamefont
  {Evans}}, \bibinfo {author} {\bibfnamefont {Y.}~\bibnamefont {Kafri}},
  \bibinfo {author} {\bibfnamefont {K.}~\bibnamefont {Sugden}}, \ and\ \bibinfo
  {author} {\bibfnamefont {J.}~\bibnamefont {Tailleur}},\ }\href
  {http://stacks.iop.org/1742-5468/2011/i=06/a=P06009} {\bibfield  {journal}
  {\bibinfo  {journal} {J. Stat. Mech.}\ }\textbf {\bibinfo {volume} {06}},\
  \bibinfo {pages} {P06009} (\bibinfo {year} {2011})}\BibitemShut {NoStop}%
\bibitem [{\citenamefont {MacDonald}\ \emph {et~al.}(1968)\citenamefont
  {MacDonald}, \citenamefont {Gibbs},\ and\ \citenamefont
  {Pipkin}}]{BIP:BIP360060102}%
  \BibitemOpen
  \bibfield  {author} {\bibinfo {author} {\bibfnamefont {C.~T.}\ \bibnamefont
  {MacDonald}}, \bibinfo {author} {\bibfnamefont {J.~H.}\ \bibnamefont
  {Gibbs}}, \ and\ \bibinfo {author} {\bibfnamefont {A.~C.}\ \bibnamefont
  {Pipkin}},\ }\href {\doibase 10.1002/bip.1968.360060102} {\bibfield
  {journal} {\bibinfo  {journal} {Biopolymers}\ }\textbf {\bibinfo {volume}
  {6}},\ \bibinfo {pages} {1} (\bibinfo {year} {1968})}\BibitemShut {NoStop}%
\bibitem [{\citenamefont {Spitzer}(1970)}]{Spitzer1970}%
  \BibitemOpen
  \bibfield  {author} {\bibinfo {author} {\bibfnamefont {F.}~\bibnamefont
  {Spitzer}},\ }\href@noop {} {\bibfield  {journal} {\bibinfo  {journal} {Adv.
  Math.}\ }\textbf {\bibinfo {volume} {5}},\ \bibinfo {pages} {246} (\bibinfo
  {year} {1970})}\BibitemShut {NoStop}%
\bibitem [{\citenamefont {Blythe}\ and\ \citenamefont
  {Evans}(2007)}]{blythe-evans:2007}%
  \BibitemOpen
  \bibfield  {author} {\bibinfo {author} {\bibfnamefont {R.~A.}\ \bibnamefont
  {Blythe}}\ and\ \bibinfo {author} {\bibfnamefont {M.~R.}\ \bibnamefont
  {Evans}},\ }\href@noop {} {\bibfield  {journal} {\bibinfo  {journal} {J.
  Phys. A}\ }\textbf {\bibinfo {volume} {40}},\ \bibinfo {pages} {R333}
  (\bibinfo {year} {2007})}\BibitemShut {NoStop}%
\bibitem [{\citenamefont {Chou}\ \emph {et~al.}(2011)\citenamefont {Chou},
  \citenamefont {Mallick},\ and\ \citenamefont
  {Zia}}]{Chou0034-4885-74-11-116601}%
  \BibitemOpen
  \bibfield  {author} {\bibinfo {author} {\bibfnamefont {T.}~\bibnamefont
  {Chou}}, \bibinfo {author} {\bibfnamefont {K.}~\bibnamefont {Mallick}}, \
  and\ \bibinfo {author} {\bibfnamefont {R.~K.~P.}\ \bibnamefont {Zia}},\
  }\href {http://stacks.iop.org/0034-4885/74/i=11/a=116601} {\bibfield
  {journal} {\bibinfo  {journal} {Rep. Prog. Phys.}\ }\textbf {\bibinfo
  {volume} {74}},\ \bibinfo {pages} {116601} (\bibinfo {year}
  {2011})}\BibitemShut {NoStop}%
\bibitem [{\citenamefont {M{\"u}ller}\ \emph {et~al.}(2005)\citenamefont
  {M{\"u}ller}, \citenamefont {Klumpp},\ and\ \citenamefont
  {Lipowsky}}]{mueller}%
  \BibitemOpen
  \bibfield  {author} {\bibinfo {author} {\bibfnamefont {M.}~\bibnamefont
  {M{\"u}ller}}, \bibinfo {author} {\bibfnamefont {S.}~\bibnamefont {Klumpp}},
  \ and\ \bibinfo {author} {\bibfnamefont {R.}~\bibnamefont {Lipowsky}},\
  }\href@noop {} {\bibfield  {journal} {\bibinfo  {journal} {J. Phys.: Condens.
  Matter}\ }\textbf {\bibinfo {volume} {17}},\ \bibinfo {pages} {S3839}
  (\bibinfo {year} {2005})}\BibitemShut {NoStop}%
\bibitem [{\citenamefont {Klumpp}\ and\ \citenamefont
  {Lipowsky}(2003)}]{klumpp}%
  \BibitemOpen
  \bibfield  {author} {\bibinfo {author} {\bibfnamefont {S.}~\bibnamefont
  {Klumpp}}\ and\ \bibinfo {author} {\bibfnamefont {R.}~\bibnamefont
  {Lipowsky}},\ }  
  {\bibfield  {journal} {\bibinfo  {journal} {J. Stat. Phys.}\ }\textbf
  {\bibinfo {volume} {113}},\ \bibinfo {pages} {233} (\bibinfo {year}
  {2003})}\BibitemShut {NoStop}%
\bibitem [{\citenamefont {Popkov}\ and\ \citenamefont
  {Sch{\"u}tz}(2003)}]{Popkov2003}%
  \BibitemOpen
  \bibfield  {author} {\bibinfo {author} {\bibfnamefont {V.}~\bibnamefont
  {Popkov}}\ and\ \bibinfo {author} {\bibfnamefont {G.~M.}\ \bibnamefont
  {Sch{\"u}tz}},\ }\href {\doibase 10.1023/A:1023819807616} {\bibfield
  {journal} {\bibinfo  {journal} {J. Stat. Phys.}\ }\textbf {\bibinfo {volume}
  {112}},\ \bibinfo {pages} {523} (\bibinfo {year} {2003})}\BibitemShut
  {NoStop}%
\bibitem [{\citenamefont {Pronina}\ and\ \citenamefont
  {Kolomeisky}(2004)}]{0305-4470-37-42-005}%
  \BibitemOpen
  \bibfield  {author} {\bibinfo {author} {\bibfnamefont {E.}~\bibnamefont
  {Pronina}}\ and\ \bibinfo {author} {\bibfnamefont {A.~B.}\ \bibnamefont
  {Kolomeisky}},\ }\href {http://stacks.iop.org/0305-4470/37/i=42/a=005}
  {\bibfield  {journal} {\bibinfo  {journal} {J. Phys. A}\ }\textbf {\bibinfo
  {volume} {37}},\ \bibinfo {pages} {9907} (\bibinfo {year}
  {2004})}\BibitemShut {NoStop}%
\bibitem [{\citenamefont {Schmittmann}\ \emph {et~al.}(2005)\citenamefont
  {Schmittmann}, \citenamefont {Krometis},\ and\ \citenamefont
  {Zia}}]{0295-5075-70-3-299}%
  \BibitemOpen
  \bibfield  {author} {\bibinfo {author} {\bibfnamefont {B.}~\bibnamefont
  {Schmittmann}}, \bibinfo {author} {\bibfnamefont {J.}~\bibnamefont
  {Krometis}}, \ and\ \bibinfo {author} {\bibfnamefont {R.~K.~P.}\ \bibnamefont
  {Zia}},\ }\href {http://stacks.iop.org/0295-5075/70/i=3/a=299} {\bibfield
  {journal} {\bibinfo  {journal} {Europhys. Lett.}\ }\textbf {\bibinfo {volume}
  {70}},\ \bibinfo {pages} {299} (\bibinfo {year} {2005})}\BibitemShut
  {NoStop}%
\bibitem [{\citenamefont {Pronina}\ and\ \citenamefont
  {Kolomeisky}(2006)}]{Pronina200612}%
  \BibitemOpen
  \bibfield  {author} {\bibinfo {author} {\bibfnamefont {E.}~\bibnamefont
  {Pronina}}\ and\ \bibinfo {author} {\bibfnamefont {A.~B.}\ \bibnamefont
  {Kolomeisky}},\ }\href {\doibase
  http://dx.doi.org/10.1016/j.physa.2006.05.006} {\bibfield  {journal}
  {\bibinfo  {journal} {Physica A}\ }\textbf {\bibinfo {volume} {372}},\
  \bibinfo {pages} {12 } (\bibinfo {year} {2006})}\BibitemShut {NoStop}%
\bibitem [{\citenamefont {Reichenbach}\ \emph {et~al.}(2007)\citenamefont
  {Reichenbach}, \citenamefont {Frey},\ and\ \citenamefont
  {Franosch}}]{1367-2630-9-6-159}%
  \BibitemOpen
  \bibfield  {author} {\bibinfo {author} {\bibfnamefont {T.}~\bibnamefont
  {Reichenbach}}, \bibinfo {author} {\bibfnamefont {E.}~\bibnamefont {Frey}}, \
  and\ \bibinfo {author} {\bibfnamefont {T.}~\bibnamefont {Franosch}},\ }\href
  {http://stacks.iop.org/1367-2630/9/i=6/a=159} {\bibfield  {journal} {\bibinfo
   {journal} {New J. Phys.}\ }\textbf {\bibinfo {volume} {9}},\ \bibinfo
  {pages} {159} (\bibinfo {year} {2007})}\BibitemShut {NoStop}%
\bibitem [{\citenamefont {Pronina}\ and\ \citenamefont
  {Kolomeisky}(2007)}]{1751-8121-40-10-004}%
  \BibitemOpen
  \bibfield  {author} {\bibinfo {author} {\bibfnamefont {E.}~\bibnamefont
  {Pronina}}\ and\ \bibinfo {author} {\bibfnamefont {A.~B.}\ \bibnamefont
  {Kolomeisky}},\ }\href {http://stacks.iop.org/1751-8121/40/i=10/a=004}
  {\bibfield  {journal} {\bibinfo  {journal} {J. Phys. A}\ }\textbf {\bibinfo
  {volume} {40}},\ \bibinfo {pages} {2275} (\bibinfo {year}
  {2007})}\BibitemShut {NoStop}%
\bibitem [{\citenamefont {Jiang}\ \emph {et~al.}(2007)\citenamefont {Jiang},
  \citenamefont {Wang},\ and\ \citenamefont {Wu}}]{Jiang2007247}%
  \BibitemOpen
  \bibfield  {author} {\bibinfo {author} {\bibfnamefont {R.}~\bibnamefont
  {Jiang}}, \bibinfo {author} {\bibfnamefont {R.}~\bibnamefont {Wang}}, \ and\
  \bibinfo {author} {\bibfnamefont {Q.-S.}\ \bibnamefont {Wu}},\ }\href
  {\doibase http://dx.doi.org/10.1016/j.physa.2006.08.025} {\bibfield
  {journal} {\bibinfo  {journal} {Physica A}\ }\textbf {\bibinfo {volume}
  {375}},\ \bibinfo {pages} {247 } (\bibinfo {year} {2007})}\BibitemShut
  {NoStop}%
\bibitem [{\citenamefont {Reichenbach}\ \emph {et~al.}(2008)\citenamefont
  {Reichenbach}, \citenamefont {Franosch},\ and\ \citenamefont
  {Frey}}]{Reichenbach2008}%
  \BibitemOpen
  \bibfield  {author} {\bibinfo {author} {\bibfnamefont {T.}~\bibnamefont
  {Reichenbach}}, \bibinfo {author} {\bibfnamefont {T.}~\bibnamefont
  {Franosch}}, \ and\ \bibinfo {author} {\bibfnamefont {E.}~\bibnamefont
  {Frey}},\ }\href {\doibase 10.1140/epje/i2008-10350-3} {\bibfield  {journal}
  {\bibinfo  {journal} {Eur. Phys. J. E}\ }\textbf {\bibinfo {volume} {27}},\
  \bibinfo {pages} {47} (\bibinfo {year} {2008})}\BibitemShut {NoStop}%
\bibitem [{\citenamefont {Wang}\ \emph {et~al.}(2008)\citenamefont {Wang},
  \citenamefont {Liu},\ and\ \citenamefont {Jiang}}]{Wang2008457}%
  \BibitemOpen
  \bibfield  {author} {\bibinfo {author} {\bibfnamefont {R.}~\bibnamefont
  {Wang}}, \bibinfo {author} {\bibfnamefont {M.}~\bibnamefont {Liu}}, \ and\
  \bibinfo {author} {\bibfnamefont {R.}~\bibnamefont {Jiang}},\ }\href
  {\doibase http://dx.doi.org/10.1016/j.physa.2007.09.042} {\bibfield
  {journal} {\bibinfo  {journal} {Physica A}\ }\textbf {\bibinfo {volume}
  {387}},\ \bibinfo {pages} {457 } (\bibinfo {year} {2008})}\BibitemShut
  {NoStop}%
\bibitem [{\citenamefont {Evans}\ \emph {et~al.}(2009)\citenamefont {Evans},
  \citenamefont {Ferrari},\ and\ \citenamefont {Mallick}}]{Evans2009}%
  \BibitemOpen
  \bibfield  {author} {\bibinfo {author} {\bibfnamefont {M.~R.}\ \bibnamefont
  {Evans}}, \bibinfo {author} {\bibfnamefont {P.~A.}\ \bibnamefont {Ferrari}},
  \ and\ \bibinfo {author} {\bibfnamefont {K.}~\bibnamefont {Mallick}},\ }\href
  {\doibase 10.1007/s10955-009-9696-2} {\bibfield  {journal} {\bibinfo
  {journal} {J. Stat. Phys.}\ }\textbf {\bibinfo {volume} {135}},\ \bibinfo
  {pages} {217} (\bibinfo {year} {2009})}\BibitemShut {NoStop}%
\bibitem [{\citenamefont {Schiffmann}\ \emph {et~al.}(2010)\citenamefont
  {Schiffmann}, \citenamefont {Appert-Rolland},\ and\ \citenamefont
  {Santen}}]{1742-5468-2010-06-P06002}%
  \BibitemOpen
  \bibfield  {author} {\bibinfo {author} {\bibfnamefont {C.}~\bibnamefont
  {Schiffmann}}, \bibinfo {author} {\bibfnamefont {C.}~\bibnamefont
  {Appert-Rolland}}, \ and\ \bibinfo {author} {\bibfnamefont {L.}~\bibnamefont
  {Santen}},\ }\href {http://stacks.iop.org/1742-5468/2010/i=06/a=P06002}
  {\bibfield  {journal} {\bibinfo  {journal} {J. Stat. Mech.}\ }\textbf
  {\bibinfo {volume} {2010}},\ \bibinfo {pages} {P06002} (\bibinfo {year}
  {2010})}\BibitemShut {NoStop}%
\bibitem [{\citenamefont {Melbinger}\ \emph {et~al.}(2011)\citenamefont
  {Melbinger}, \citenamefont {Reichenbach}, \citenamefont {Franosch},\ and\
  \citenamefont {Frey}}]{PhysRevE.83.031923}%
  \BibitemOpen
  \bibfield  {author} {\bibinfo {author} {\bibfnamefont {A.}~\bibnamefont
  {Melbinger}}, \bibinfo {author} {\bibfnamefont {T.}~\bibnamefont
  {Reichenbach}}, \bibinfo {author} {\bibfnamefont {T.}~\bibnamefont
  {Franosch}}, \ and\ \bibinfo {author} {\bibfnamefont {E.}~\bibnamefont
  {Frey}},\ }\href {\doibase 10.1103/PhysRevE.83.031923} {\bibfield  {journal}
  {\bibinfo  {journal} {Phys. Rev. E}\ }\textbf {\bibinfo {volume} {83}},\
  \bibinfo {pages} {031923} (\bibinfo {year} {2011})}\BibitemShut {NoStop}%
\bibitem [{\citenamefont {Saha}\ and\ \citenamefont {Mukherji}(2013)}]{saha}%
  \BibitemOpen
  \bibfield  {author} {\bibinfo {author} {\bibfnamefont {B.}~\bibnamefont
  {Saha}}\ and\ \bibinfo {author} {\bibfnamefont {S.}~\bibnamefont
  {Mukherji}},\ }\href {http://stacks.iop.org/1742-5468/2013/i=09/a=P09004}
  {\bibfield  {journal} {\bibinfo  {journal} {J. Stat. Mech.}\ }\textbf
  {\bibinfo {volume} {09}},\ \bibinfo {pages} {P09004} (\bibinfo {year}
  {2013})}\BibitemShut {NoStop}%
\bibitem [{\citenamefont {Gupta}\ and\ \citenamefont
  {Dhiman}(2014)}]{PhysRevE.89.022131}%
  \BibitemOpen
  \bibfield  {author} {\bibinfo {author} {\bibfnamefont {A.~K.}\ \bibnamefont
  {Gupta}}\ and\ \bibinfo {author} {\bibfnamefont {I.}~\bibnamefont {Dhiman}},\
  }\href {\doibase 10.1103/PhysRevE.89.022131} {\bibfield  {journal} {\bibinfo
  {journal} {Phys. Rev. E}\ }\textbf {\bibinfo {volume} {89}},\ \bibinfo
  {pages} {022131} (\bibinfo {year} {2014})}\BibitemShut {NoStop}%
\bibitem [{\citenamefont {Johann}\ \emph {et~al.}(2014)\citenamefont {Johann},
  \citenamefont {Goswami},\ and\ \citenamefont {Kruse}}]{PhysRevE.89.042713}%
  \BibitemOpen
  \bibfield  {author} {\bibinfo {author} {\bibfnamefont {D.}~\bibnamefont
  {Johann}}, \bibinfo {author} {\bibfnamefont {D.}~\bibnamefont {Goswami}}, \
  and\ \bibinfo {author} {\bibfnamefont {K.}~\bibnamefont {Kruse}},\ }\href
  {\doibase 10.1103/PhysRevE.89.042713} {\bibfield  {journal} {\bibinfo
  {journal} {Phys. Rev. E}\ }\textbf {\bibinfo {volume} {89}},\ \bibinfo
  {pages} {042713} (\bibinfo {year} {2014})}\BibitemShut {NoStop}%
\bibitem [{\citenamefont {Pinkoviezky}\ and\ \citenamefont
  {Gov}(2014)}]{pinkoviezky}%
  \BibitemOpen
  \bibfield  {author} {\bibinfo {author} {\bibfnamefont {I.}~\bibnamefont
  {Pinkoviezky}}\ and\ \bibinfo {author} {\bibfnamefont {N.~S.}\ \bibnamefont
  {Gov}},\ }\href {\doibase 10.1103/PhysRevE.89.052703} {\bibfield  {journal}
  {\bibinfo  {journal} {Phys. Rev. E}\ }\textbf {\bibinfo {volume} {89}},\
  \bibinfo {pages} {052703} (\bibinfo {year} {2014})}\BibitemShut {NoStop}%
\bibitem [{\citenamefont {Curatolo}\ \emph {et~al.}(2016)\citenamefont
  {Curatolo}, \citenamefont {Evans}, \citenamefont {Kafri},\ and\ \citenamefont
  {Tailleur}}]{1751-8121-49-9-095601}%
  \BibitemOpen
  \bibfield  {author} {\bibinfo {author} {\bibfnamefont {A.~I.}\ \bibnamefont
  {Curatolo}}, \bibinfo {author} {\bibfnamefont {M.~R.}\ \bibnamefont {Evans}},
  \bibinfo {author} {\bibfnamefont {Y.}~\bibnamefont {Kafri}}, \ and\ \bibinfo
  {author} {\bibfnamefont {J.}~\bibnamefont {Tailleur}},\ }\href
  {http://stacks.iop.org/1751-8121/49/i=9/a=095601} {\bibfield  {journal}
  {\bibinfo  {journal} {J. Phys. A}\ }\textbf {\bibinfo {volume} {49}},\
  \bibinfo {pages} {095601} (\bibinfo {year} {2016})}\BibitemShut {NoStop}%
\bibitem [{Sup()}]{Supplement}%
  \BibitemOpen
  \bibinfo {note} {See Supplemental Material at [url
  will be inserted by publisher].} \BibitemShut {Stop}%
\bibitem [{\citenamefont {Reese}\ \emph {et~al.}(2014)\citenamefont {Reese},
  \citenamefont {Melbinger},\ and\ \citenamefont {Frey}}]{Reese20140031}%
  \BibitemOpen
  \bibfield  {author} {\bibinfo {author} {\bibfnamefont {L.}~\bibnamefont
  {Reese}}, \bibinfo {author} {\bibfnamefont {A.}~\bibnamefont {Melbinger}}, \
  and\ \bibinfo {author} {\bibfnamefont {E.}~\bibnamefont {Frey}},\ }\href
  {\doibase 10.1098/rsfs.2014.0031} {\bibfield  {journal} {\bibinfo  {journal}
  {Interface Focus}\ }\textbf {\bibinfo {volume} {4}} (\bibinfo {year}
  {2014}),\ 10.1098/rsfs.2014.0031}\BibitemShut {NoStop}%
\bibitem [{\citenamefont {Klumpp}\ \emph {et~al.}(2007)\citenamefont {Klumpp},
  \citenamefont {M{\"u}ller},\ and\ \citenamefont {Lipowsky}}]{Klumpp2007}%
  \BibitemOpen
  \bibfield  {author} {\bibinfo {author} {\bibfnamefont {S.}~\bibnamefont
  {Klumpp}}, \bibinfo {author} {\bibfnamefont {M.~J.~I.}\ \bibnamefont
  {M{\"u}ller}}, \ and\ \bibinfo {author} {\bibfnamefont {R.}~\bibnamefont
  {Lipowsky}},\ }\enquote {\bibinfo {title} {Traffic and Granular Flow '05}}\ \
  (\bibinfo  {publisher} {Springer},\ \bibinfo {address}
  {Berlin},\ \bibinfo {year} {2007})
  \BibitemShut
  {NoStop}%
\bibitem [{\citenamefont {Gillespie}(1976)}]{GILLESPIE1976403}%
  \BibitemOpen
  \bibfield  {author} {\bibinfo {author} {\bibfnamefont {D.~T.}\ \bibnamefont
  {Gillespie}},\ }\href {\doibase
  http://dx.doi.org/10.1016/0021-9991(76)90041-3} {\bibfield  {journal}
  {\bibinfo  {journal} {J. Comput. Phys.}\ }\textbf {\bibinfo {volume} {22}},\
  \bibinfo {pages} {403 } (\bibinfo {year} {1976})}\BibitemShut {NoStop}%
\bibitem [{\citenamefont {Derrida}\ \emph {et~al.}(1995)\citenamefont
  {Derrida}, \citenamefont {Evans},\ and\ \citenamefont
  {Mallick}}]{Derrida1995}%
  \BibitemOpen
  \bibfield  {author} {\bibinfo {author} {\bibfnamefont {B.}~\bibnamefont
  {Derrida}}, \bibinfo {author} {\bibfnamefont {M.~R.}\ \bibnamefont {Evans}},
  \ and\ \bibinfo {author} {\bibfnamefont {K.}~\bibnamefont {Mallick}},\ }\href
  {\doibase 10.1007/BF02181206} {\bibfield  {journal} {\bibinfo  {journal} {J.
  Stat. Phys.}\ }\textbf {\bibinfo {volume} {79}},\ \bibinfo {pages} {833}
  (\bibinfo {year} {1995})}\BibitemShut {NoStop}%
\bibitem [{\citenamefont {Kolomeisky}\ \emph {et~al.}(1998)\citenamefont
  {Kolomeisky}, \citenamefont {Sch{\"u}tz}, \citenamefont {Kolomeisky},\ and\
  \citenamefont {Straley}}]{0305-4470-31-33-003}%
  \BibitemOpen
  \bibfield  {author} {\bibinfo {author} {\bibfnamefont {A.~B.}\ \bibnamefont
  {Kolomeisky}}, \bibinfo {author} {\bibfnamefont {G.~M.}\ \bibnamefont
  {Sch{\"u}tz}}, \bibinfo {author} {\bibfnamefont {E.~B.}\ \bibnamefont
  {Kolomeisky}}, \ and\ \bibinfo {author} {\bibfnamefont {J.~P.}\ \bibnamefont
  {Straley}},\ }\href {http://stacks.iop.org/0305-4470/31/i=33/a=003}
  {\bibfield  {journal} {\bibinfo  {journal} {J. Phys. A}\ }\textbf {\bibinfo
  {volume} {31}},\ \bibinfo {pages} {6911} (\bibinfo {year}
  {1998})}\BibitemShut {NoStop}%
\bibitem [{\citenamefont {Popkov}\ \emph {et~al.}(2003)\citenamefont {Popkov},
  \citenamefont {R\'akos}, \citenamefont {Willmann}, \citenamefont
  {Kolomeisky},\ and\ \citenamefont {Sch\"utz}}]{PhysRevE.67.066117}%
  \BibitemOpen
  \bibfield  {author} {\bibinfo {author} {\bibfnamefont {V.}~\bibnamefont
  {Popkov}}, \bibinfo {author} {\bibfnamefont {A.}~\bibnamefont {R\'akos}},
  \bibinfo {author} {\bibfnamefont {R.~D.}\ \bibnamefont {Willmann}}, \bibinfo
  {author} {\bibfnamefont {A.~B.}\ \bibnamefont {Kolomeisky}}, \ and\ \bibinfo
  {author} {\bibfnamefont {G.~M.}\ \bibnamefont {Sch\"utz}},\ }\href {\doibase
  10.1103/PhysRevE.67.066117} {\bibfield  {journal} {\bibinfo  {journal} {Phys.
  Rev. E}\ }\textbf {\bibinfo {volume} {67}},\ \bibinfo {pages} {066117}
  (\bibinfo {year} {2003})}\BibitemShut {NoStop}%
\bibitem [{\citenamefont {Tsekouras}\ and\ \citenamefont
  {Kolomeisky}(2008)}]{1751-8121-41-9-095002}%
  \BibitemOpen
  \bibfield  {author} {\bibinfo {author} {\bibfnamefont {K.}~\bibnamefont
  {Tsekouras}}\ and\ \bibinfo {author} {\bibfnamefont {A.~B.}\ \bibnamefont
  {Kolomeisky}},\ }\href {http://stacks.iop.org/1751-8121/41/i=9/a=095002}
  {\bibfield  {journal} {\bibinfo  {journal} {J.  Phys. A}\ }\textbf {\bibinfo {volume} {41}},\ \bibinfo
  {pages} {095002} (\bibinfo {year} {2008})}\BibitemShut {NoStop}%
\bibitem [{\citenamefont {Reithmann}\ \emph {et~al.}(2016)\citenamefont
  {Reithmann}, \citenamefont {Reese},\ and\ \citenamefont
  {Frey}}]{PhysRevLett.117.078102}%
  \BibitemOpen
  \bibfield  {author} {\bibinfo {author} {\bibfnamefont {E.}~\bibnamefont
  {Reithmann}}, \bibinfo {author} {\bibfnamefont {L.}~\bibnamefont {Reese}}, \
  and\ \bibinfo {author} {\bibfnamefont {E.}~\bibnamefont {Frey}},\ }\href
  {\doibase 10.1103/PhysRevLett.117.078102} {\bibfield  {journal} {\bibinfo
  {journal} {Phys. Rev. Lett.}\ }\textbf {\bibinfo {volume} {117}},\ \bibinfo
  {pages} {078102} (\bibinfo {year} {2016})}\BibitemShut {NoStop}%
\end{thebibliography}
\end{document}